\documentclass[a4paper,10pt]{article}
\usepackage[utf8]{inputenc}
\usepackage{amsthm}
\usepackage{amsfonts}
\usepackage{amssymb}	
\usepackage{amsmath}
\allowdisplaybreaks
\usepackage{mathtools}
\usepackage[english]{babel}
\usepackage{color}
\usepackage{slashed}
\usepackage{enumerate}
\usepackage{graphicx}
\usepackage{systeme}
\usepackage{dsfont}
\usepackage{relsize}
\usepackage[multiple]{footmisc}
\usepackage[margin=1in]{geometry}
\usepackage{float}
\usepackage{tikz}
\usepackage[labelformat=simple]{subcaption}

\usepackage{graphicx}
\usepackage{bmpsize}
\usepackage{epstopdf}
\usepackage{bbm}
\usepackage{xcolor,etoolbox}
\usepackage{titling}
\usepackage{authblk}
\newcommand*\samethanks[1][\value{footnote}]{\footnotemark[#1]}
\usepackage{blindtext,graphicx}
\usepackage[absolute]{textpos}
\usepackage{physics}
\usepackage[giveninits=true,sortcites=true,date=year,maxbibnames=99,doi=false,isbn=false,url=false,eprint=false]{biblatex}
\renewbibmacro{in:}{}
\DeclareFieldFormat{pages}{#1}
\AtEveryBibitem{%
  \clearlist{language}%
}
\usepackage{csquotes}

\usepackage{setspace}
\theoremstyle{definition}

\title{\vspace{-2cm}Analytical solutions of the simple shear problem for certain types of micromorphic continuum models – including full derivations}
\author{Gianluca Rizzi\thanks{GEOMAS, INSA-Lyon, Universit\'e de Lyon, 20 avenue Albert Einstein,	69621, Villeurbanne cedex, France} \quad and \quad Geralf Hütter\thanks{TU Bergakademie Freiberg, Institute of Mechanics and Fluid Dynamics, Lampadiusstr. 4, 09596 Freiberg, Germany} \quad and \quad Angela Madeo\samethanks[1] \quad and \quad Patrizio Neff\thanks{Head of Chair for Nonlinear Analysis and Modelling, Fakultät für Mathematik, Universität Duisburg-Essen, \\ \indent Thea-Leymann-Straße 9, 45127 Essen, Germany}}

\thanksmarkseries{arabic}

\date{\today}
\begin{document}

	\maketitle
	\small
	\begin{abstract}
		To draw conclusions as regards the stability and modelling limits of the investigated continuum, we consider a family of infinitesimal isotropic generalized continuum models (Mindlin-Eringen micromorphic, relaxed micromorphic continuum, Cosserat, micropolar, microstretch, microstrain, microvoid, indeterminate couple stress, second gradient elasticity, etc.) and solve analytically the simple shear problem of an infinite stripe. A qualitative measure characterizing the different generalized continuum moduli is given by the shear stiffness $\mu^{*}$. This stiffness is in general length-scale dependent. Interesting limit cases are highlighted, which allow to interpret some of the appearing material parameter of the investigated continua.
	\end{abstract}
	
\section{Introduction}
\label{sec:intro}
Today there exist a huge variety of small strain, linear generalized continuum models that allow to extend the modelling capabilities to include size dependent response.
We mention Mindlin-Eringen micromorphic \cite{eringen1968mechanics,mindlin1964micro,Forest2018,Forest2013_CISM}, relaxed micromorphic continuum \cite{neff2014unifying,neff2019identification,d2019effective}, Cosserat \cite{cosserat1909theorie,neff2006cosserat,neff2009new,neff2010stable,neff2009simple,jeong2008existence}, micropolar, microstretch \cite{mindlin1964micro,neff2014unifying}, microstrain \cite{forest2006nonlinear}, microvoid \cite{neff2014unifying}, indeterminate couple stress \cite{neff2009simple,munch2017modified,madeo2016new,neff2016some,hadjesfandiari2011couple}, second gradient elasticity \cite{mindlin1964micro,neff2010stable}, etc.

The basic problem of all these theories, even for the infinitesimal strain isotropic case under consideration here, is the huge number of newly appearing constitutive coefficient which need to be determined and physically interpreted.
Homogeneous tests fail to reveal the inherent size-effects and are therefore not sufficient to determine and interpret those constitutive coefficients, that are connected to these size-effects.
To gain further insight, it is therefore mandatory to investigate boundary value problems which produce some inhomogenes response (as it can be seen in real or virtual experiments \cite{rueger2019experimental,dunn2020size,yoder2019size,nourmohammadi2020effective,rizzi2019identification,pham2020influence}).
There exist a couple of inhomogeneous analytical solutions to simple shear \cite{rizzi2019identification,zhang2011analytical,hutter2019micro,aifantis2005role,aifantis1987physics,diebels2003stress,forest2013questioning,iltchev2015computational,kruch1998computation,liebenstein2018size,maziere2015strain,tekouglu2008size}, pure bending and torsion for some of the simpler models mentioned above.
\begin{figure}[H]
	\centering
	\begin{tikzpicture}
	\coordinate (A) at (-2.5,0);
	\coordinate (A1) at (-0.5,0);
	\coordinate (A2) at (-1,0);
	\coordinate (B) at (2.5,0);
	\coordinate (B1) at (6.5,0);
	\coordinate (B2) at (7,0);
	\coordinate (C) at (-2.5,2);
	\coordinate (C1) at (1,2);
	\coordinate (C2) at (-0.5,2);
	\coordinate (C3) at (-1,2);
	\coordinate (D) at (2.5,2);
	\coordinate (D1) at (6,2);
	\coordinate (D2) at (6.5,2);
	\coordinate (D3) at (7,2);
	\draw[thick] (A1) -- (B1);
	\draw[thick] (C2) -- (D2);
	\draw[thick,dashed] (A1) -- (A2);
	\draw[thick,dashed] (C2) -- (C3);
	\draw[thick,dashed] (B1) -- (B2);
	\draw[thick,dashed] (D2) -- (D3);
	\draw[thick,dashed] (0.0,0) -- (0.0,2);
	\draw[thick,dashed] (5.0,0) -- (5.0,2);
	\draw[thick] plot[domain = 0:2] ({(\x*cosh(1)+sinh(1-\x))*exp(1)/2-exp(1)/2*sinh(1)},\x);
	\draw[thick] plot[domain = 0:2] ({(\x*cosh(1)+sinh(1-\x))*exp(1)/2-exp(1)/2*sinh(1)+5},\x);
	\draw[thick,->] (7,0) -- (7,0.5);
	\node[left] at (7,0.25) {$x_2$};
	\draw[thick,->] (7,0) -- (7.5,0);
	\node[left] at (7.5,-0.25) {$x_1$};
	\draw[-latex] (0,2.25) -- (0.9,2.25);	
	\draw[-latex] (1,2.25) -- (1.9,2.25);
	\draw[-latex] (2,2.25) -- (2.9,2.25);	
	\draw[-latex] (3,2.25) -- (3.9,2.25);
	\draw[-latex] (4,2.25) -- (4.9,2.25);
	\draw[-latex] (5,2.25) -- (5.9,2.25);		
	\node[above] at (3,2.25) {$\boldsymbol{\gamma} \, h$};
	\draw[-] (-0.5,-0.15) -- (-0.1,-0.);		
	\draw[-] (0,-0.15) -- (0.4,-0.);
	\draw[-] (0.5,-0.15) -- (0.9,-0.);	
	\draw[-] (1,-0.15) -- (1.4,-0.);
	\draw[-] (1.5,-0.15) -- (1.9,-0.);
	\draw[-] (2,-0.15) -- (2.4,-0.);
	\draw[-] (2.5,-0.15) -- (2.9,-0.);
	\draw[-] (3,-0.15) -- (3.4,-0.);
	\draw[-] (3.5,-0.15) -- (3.9,-0.);
	\draw[-] (4,-0.15) -- (4.4,-0.);
	\draw[-] (4.5,-0.15) -- (4.9,-0.);
	\draw[-] (5,-0.15) -- (5.4,-0.);
	\draw[-] (5.5,-0.15) -- (5.9,-0.);
	\draw[-] (6,-0.15) -- (6.4,-0.);
	\draw[latex-latex] (-1,0) -- (-1,2);	
	\node[right] at (-1.5,1) {$h$};
	\end{tikzpicture}
	\caption{Scketch of an infinite stripe subjected to simple shear boundary conditions.}
	\label{fig:intro}
\end{figure}
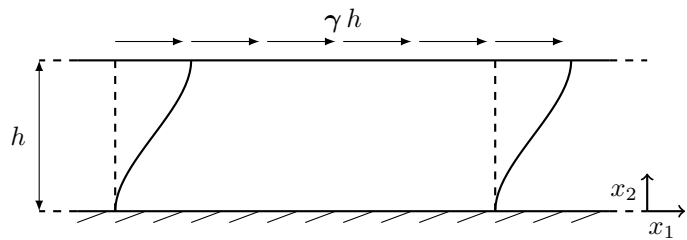
Cosserat theory has widely been used to model such size effects since respective analytical solutions are available (and Cosserat has less parameters of course) whereas few analytical solutions are available for more general classes of micromorphic theories (except those which are already cited) and that such solutions are necessary to promote micromorphic theories.

In this paper, we focus on the analytical solution for the simple shear of an infinite stripe, which we provide for a family of generalized continua.
The method to obtain these solution of linear problems is fairly standard, but still needs a concentrated effort.
The deep interest occurs in comparing the resulting size-dependent shear stiffness 
$$\mu^{*} = \dfrac{\widetilde{\sigma}_{12}}{\boldsymbol{\gamma}}$$
where $\widetilde{\boldsymbol{\sigma}}$ is force-stress tensor and $\boldsymbol{\gamma}$ is the shear deformation for the classical shear problem.
The factor $\mu^{*}$ will, in general, depend on a length-scale parameter ($L_{\text{c}}\geq0$), the height $h$ of the stripe (Fig.~\ref{fig:intro}), and on the other parameters of the generalized continua.
It is clear that for material samples with infinite relative width (corresponding to $L_{\text{c}} \to 0$, viz $h \to \infty$), we need to recover the classical size-independent shear modulus $\mu^{*} \equiv \mu _{\text{macro}}$.
This imposes already some telling relations between the remaining parameters of the model.

Another interesting limit case concerns the shear stiffness for arbitrary ``flat" samples ($h \to 0$), which correspond to $L_{\text{c}} \to \infty$.
Then, typically, $\eval{\mu^{*}}_{L_{\text{c}} \to \infty}$ is an increasing function of $L_{\text{c}}$.
Whether or not there exists an upper bound on the shear stiffness depends then on the specific model.
Typically, gradient elasticity (including the indeterminate couple stress model) exhibits an unphysical stiffness singularity.
Other more general models may have the same stiffness singularity in conjunction with the limit to infinity for further material parameters.

As a first result we can note that the simple shear of an infinite block is triggering an inhomogeneous solution when forced by the boundary condition, but this inhomogeneity remains for many of the investigated models ``tame'', in the sense that the shear stiffness remains bounded as $L_{\text{c}} \to \infty$ ($h \to 0$).
We expect this to change completely when we repeat this kind of investigation for the pure bending case in a future contribution.

Since we aim at a readable exposition we use direct tensor notation alongside index-notation.
Moreover, due to the large number of constitutive coefficients, especially in the curvature  energies of the different models, we have opted to simplify the curvature expressions to a 1-parameter format for the classical micromorphic model, the microstrain model and the gradient elasticity model.
For all other models, the effective curvature parameters are completely taken into account, nevertheless, they simplify considerably for the chosen 2D simple shear problem.

All our investigated generalized continuum models are in one way or another derived from the general Mindlin-Eringen \cite{mindlin1964micro} micromorphic framework.
They differ in the choice of the additional independent degrees of freedom (Cosserat: 3 rotation dof, microstrain: 6 strain dof, etc.) besides the classical 3 translation degrees of freedom.
They also differ in the choice of curvature parameters. The different curvature terms also presuppose according different boundary conditions, e.g. if the full gradient of the micro distortion $\boldsymbol{P}$ is used as curvature (as in the classical micromorphic model) $\left\lVert \boldsymbol{\nabla P} \right\rVert^2$, then one can  impose Dirichlet boundary conditions on $\boldsymbol{P}$ completely.
If, on the other hand, only the Curl of the micro distortion $\boldsymbol{P}$ is controlled, as in the relaxed micromorphic model $\left\lVert \mbox{Curl} \boldsymbol{P} \right\rVert^2$, then only tangential boundary conditions on $\boldsymbol{P}$ can be prescribed.

In second gradient elasticity, one controls $\left\lVert \boldsymbol{\nabla^2 u} \right\rVert^2$ and accordingly, the values of $\boldsymbol{\nabla u}$ at the boundary can be given, similarly for the indeterminate-couple stress model, in which $\left\lVert \boldsymbol{\nabla} \mbox{curl} \, \boldsymbol{u} \right\rVert^2$ is controlled and one may prescribe the values of $\mbox{curl} \, \boldsymbol{u}$ at the boundary.
For the Cosserat model, the curvature measure can be taken to be $\left\lVert \mbox{Curl} \, \boldsymbol{A} \right\rVert^2$, for a skew-symmetric matrix $\boldsymbol{A}$.
Since Curl controls all first partial derivatives of $\boldsymbol{A}$ the tangential condition for skew-symmetric A is equivalent to a full Dirichlet condition.

The apparent size-effect inherent in all these theories is strongly related to the employed Dirichlet boundary conditions for the additional (or higher order) terms. Indeed, if the additional degrees of freedom are left free at the upper and lower surface (for simple shear), then the solution turns into the homogeneous classical simple shear solution without size-effects.

\subsection{Notation}
For vectors $\boldsymbol{a},\boldsymbol{b}\in\mathbb{R}^n$, we consider the scalar product  $\langle \boldsymbol{a},\boldsymbol{b} \rangle \coloneqq \sum_{i=1}^n a_i\,b_i \in \mathbb{R}$, the (squared) norm  $\norm{\boldsymbol{a}}^2\coloneqq\langle \boldsymbol{a},\boldsymbol{a} \rangle$ and  the dyadic product  $\boldsymbol{a}\otimes \boldsymbol{b} \coloneqq \left(a_i\,b_j\right)_{i,j=1,\ldots,n}\in \mathbb{R}^{n\times n}$. Similarly, for tensors  $\boldsymbol{P},\boldsymbol{Q}\in\mathbb{R}^{n\times n}$  with Cartesian coordinates $P_{ij}$ and $Q_{ij}$, we define the scalar product $\langle \boldsymbol{P},\boldsymbol{Q} \rangle \coloneqq\sum_{i,j=1}^n P_{ij}\,Q_{ij} \in \mathbb{R}$ and the (squared) Frobenius-norm $\norm{\boldsymbol{P}}^2\coloneqq\langle \boldsymbol{P},\boldsymbol{P} \rangle$.
Moreover, $\boldsymbol{P}^T\coloneqq (P_{ji})_{i,j=1,\ldots,n}$ denotes the transposition of the matrix $\boldsymbol{P}=(P_{ij})_{i,j=1,\ldots,n}$, which decomposes orthogonally into the symmetric part $\mbox{sym} \boldsymbol{P} \coloneqq \frac{1}{2} (\boldsymbol{P}+\boldsymbol{P}^T)$ and the skew-symmetric part $\mbox{skew} \boldsymbol{P} \coloneqq \frac{1}{2} (\boldsymbol{P}-\boldsymbol{P}^T )$.
The Lie-Algebra of skew-symmetric matrices is denoted by $\mathfrak{so}(3)\coloneqq \{\boldsymbol{A}\in\mathbb{R}^{3\times 3}\mid \boldsymbol{A}^T = -\boldsymbol{A}\}$.
The identity matrix is denoted by $\boldsymbol{\mathbbm{1}}$, so that the trace of a matrix $\boldsymbol{P}$ is given by \ $\tr \boldsymbol{P} \coloneqq \langle \boldsymbol{P},\boldsymbol{\mathbbm{1}} \rangle$.
%
Using the bijection $\mbox{axl}:\mathfrak{so}(3)\to\mathbb{R}^3$ we have
\begin{equation}
\boldsymbol{A} \, \boldsymbol{b} =\mbox{axl}(\boldsymbol{A})\times \boldsymbol{b} \quad \forall\, \boldsymbol{A}\in\mathfrak{so}(3), \boldsymbol{b}\in\mathbb{R}^3.
\end{equation}
where $\times$ denotes the standard cross product in $\mathbb{R}^3$.
The gradient and the curl for a vector field $\boldsymbol{u}$ are defined as
\begin{equation}
\boldsymbol{\nabla u}
=\!
\left(
\begin{array}{ccc}
u_{1,1} & u_{1,2} & u_{1,3} \\
u_{2,1} & u_{2,2} & u_{2,3} \\
u_{3,1} & u_{3,2} & u_{3,3}
\end{array}
\right)\, ,
\quad
\mbox{curl} \, \boldsymbol{u} = \boldsymbol{\nabla} \times \boldsymbol{u}
=
\left(
\begin{array}{ccc}
u_{3,2} - u_{2,3}  \\
u_{1,3} - u_{3,1}  \\
u_{2,1} - u_{1,2}
\end{array}
\right)
 \, .
\end{equation}
Moreover, we introduce the $\mbox{Curl}$, the $\mbox{Div}$ and the gradient operators of the matrix $\boldsymbol{P}$ as:
\begin{equation}
\mbox{Curl} \, \boldsymbol{P}
=\!
\left(
\begin{array}{c}
\mbox{curl}\left( P_{11} \, , \right. \, P_{12} \, , \, \left. P_{13} \right) \\
\mbox{curl}\left( P_{21} \, , \right. \, P_{22} \, , \, \left. P_{23} \right) \\
\mbox{curl}\left( P_{31} \, , \right. \, P_{32} \, , \, \left. P_{33} \right) 
\end{array}
\right) \!,
\quad
\mbox{Div}  \, \boldsymbol{P}
=\!
\left(
\begin{array}{c}
\mbox{div}\left( P_{11} \, , \right. \, P_{12} \, , \, \left. P_{13} \right) \\
\mbox{div}\left( P_{21} \, , \right. \, P_{22} \, , \, \left. P_{23} \right) \\
\mbox{div}\left( P_{31} \, , \right. \, P_{32} \, , \, \left. P_{33} \right) 
\end{array}
\right),
\end{equation}

\section{Simple shear for the isotropic Cauchy continuum}
The expression of the strain energy for an isotropic Cauchy continuum is
\begin{equation}
W \left(\boldsymbol{\nabla u}\right) = 
\mu _{\text{macro}} \left\lVert \mbox{sym} \boldsymbol{\nabla u} \right\rVert^{2} + 
\dfrac{\lambda_{\text{macro}}}{2} \mbox{tr}^2\left(\boldsymbol{\nabla u}\right)
\label{eq:energyCau}
\end{equation}
while the equilibrium equations without body forces are
\begin{equation}
\mbox{Div}\left[ 
2\,\mu _{\text{macro}}\,\mbox{sym} \boldsymbol{\nabla u}
+ \lambda_{\text{macro}}\,\mbox{tr}\left(\boldsymbol{\nabla u}\right)  \boldsymbol{\mathbbm{1}} 
\right] 
= \boldsymbol{0}.
\label{eq:equiCau}
\end{equation}
The boundary conditions for the simple shear problem are $u_{1}(x_2=0) = 0$, $u_{1}(x_2=h) = \boldsymbol{\gamma} \, h$, $u_{2} (x_2=0) = 0$, and $u_{2} (x_2=h) = 0$ (see Fig.~\ref{fig:intro}).
The displacement fields solution and the shear stiffness for the simple shear problem are the following:
\begin{equation}
u_{1} (x_2) = \boldsymbol{\gamma} \, x_{2} \, ,
\quad u_{2} (x_2) = 0 \, ,
\quad
\mu^{*} = \dfrac{\sigma_{12}}{\boldsymbol{\gamma}} = \mu _{\text{macro}}.
\end{equation}

Here and in the remainder of this work, the elastic coefficients $\mu_i,\lambda_i$ are expressed in [MPa], the shear deformation $\boldsymbol{\gamma}$ is dimensionless, the lengths $L_{\text{c}}$ and the thickness $h$ in meter [m].

\section{Classical Mindlin-Eringen formulation}
The classical micromorphic model couples the displacement $\boldsymbol{u} \in \mathbb{R}^{3}$ with an affine field $\boldsymbol{P} \in \mathbb{R}^{3\times3}$, called the micro-distortion. In the isotropic case, the elastic energy can be represented as 
\begin{equation*}
\begin{split}
	W \left(\boldsymbol{\nabla u}, \boldsymbol{P}, \boldsymbol{\nabla P}\right) = & \,
		\widehat{\mu} \, \left \lVert \mbox{sym} \, \boldsymbol{\nabla u} \right \rVert ^2
		+ \dfrac{\widehat{\lambda}}{2} \, \mbox{tr}^2 \left(\boldsymbol{\nabla u}\right)
		+ \dfrac{b_1}{2} \, \mbox{tr}^2 \left(\boldsymbol{\nabla u} - \boldsymbol{P}\right)
		\\[3mm]
		&
		+ \dfrac{b_2}{2} \, \left \lVert \boldsymbol{\nabla u} - \boldsymbol{P} \right \rVert ^2
		+ \dfrac{b_3}{2} \, \langle \boldsymbol{\nabla u} - \boldsymbol{P},\left(\boldsymbol{\nabla u} - \boldsymbol{P}\right)^{T} \rangle
		+ g_1 \, \mbox{tr} \left(\boldsymbol{\nabla u}\right)\mbox{tr} \left(\boldsymbol{\nabla u} - \boldsymbol{P}\right)  \\[3mm]
		&
		+ g_2 \, \langle \mbox{sym} \, \boldsymbol{\nabla u},\left(\boldsymbol{\nabla u} - \boldsymbol{P}\right)^{T} \rangle
		+ \dfrac{1}{2} \langle \mathbb{A}\boldsymbol{\nabla P},\boldsymbol{\nabla P} \rangle =
		\\[3mm] 
\end{split}
\end{equation*}
\begin{equation}
\begin{split}
\qquad		&
		\!\!\!\!\!\!\!
		=\, \widehat{\mu} \, \epsilon_{ij} \, \epsilon_{ij} 
		+ \dfrac{\widehat{\lambda}}{2} \, \epsilon_{ii} \, \epsilon_{jj}
		+ \dfrac{b_1}{2} \, \boldsymbol{\gamma}_{ii} \, \boldsymbol{\gamma}_{jj} 
		+ \dfrac{b_2}{2} \, \boldsymbol{\gamma}_{ij} \, \boldsymbol{\gamma}_{ij} 
		+ \dfrac{b_3}{2} \, \boldsymbol{\gamma}_{ij} \, \boldsymbol{\gamma}_{ji}
		\\[3mm]
		&
		+ g_1 \, \boldsymbol{\gamma}_{ii} \, \epsilon_{jj} 
		+ g_2 \, \left(\boldsymbol{\gamma}_{ij} + \boldsymbol{\gamma}_{ji} \, \right)\epsilon_{ij}
		+ a_1 \, \chi_{iik} \, \chi_{kjj} 
		+ a_2 \, \chi_{iik} \, \chi_{jkj}
		\\[3mm]
		&
		+ \dfrac{1}{2} \, a_3 \, \chi_{iik} \, \chi_{jjk} 
		+ \dfrac{1}{2} \, a_4 \, \chi_{ijj} \, \chi_{ikk} 
		+ a_5 \, \chi_{ijj} \, \chi_{kik}
		\\[3mm]
		&
		+ \dfrac{1}{2} \, a_8 \, \chi_{iji} \, \chi_{kjk}
		+ \dfrac{1}{2} \, a_{10} \, \chi_{ijk} \, \chi_{ijk} 
		+ a_{11} \, \chi_{ijk} \, \chi_{jki}
		\\[3mm]
		&
		+ \dfrac{1}{2} \, a_{13} \, \chi_{ijk} \, \chi_{ikj} 
		+ \dfrac{1}{2} \, a_{14} \, \chi_{ijk} \, \chi_{jik}
		+ \dfrac{1}{2} \, a_{15} \, \chi_{ijk} \, \chi_{kji}
\end{split}
\label{eq:energy_MM_Mind}
\end{equation}
where $\boldsymbol{\epsilon} = \mbox{sym} \, \boldsymbol{\nabla u}$ is the symmetric part of the gradient of the displacement field, $\boldsymbol{\boldsymbol{\gamma}} = \boldsymbol{\nabla u} - \boldsymbol{P}$ is the difference between the gradient of the displacement field and the micro-distortion tensor, and $\chi_{ijk} = P_{jk,i}$ is the gradient of the micro-distortion.

To the authors knowledge, the only simple shear analytical solution available in the literature for this model is obtained for a very restrictive choice of parameters \cite{zhang2011analytical,hutter2015micromorphic}
\footnote{
	The following energy expression has been used in \cite{zhang2011analytical}: 
	$W \left(\boldsymbol{\nabla u}, \boldsymbol{P}, \boldsymbol{\nabla P}\right) = 
	  \mu \left\lVert \mbox{sym} \, \boldsymbol{\nabla u} \right\rVert^2
	+ \lambda/2 \, \mbox{tr}^2 \left(\boldsymbol{\nabla u} \right)
	+ \alpha \, \mu \left\lVert \boldsymbol{\nabla u} - \boldsymbol{P} \right\rVert^2
	+ \alpha \, \lambda/2 \, \mbox{tr}^2 \left( \boldsymbol{\nabla u}  - \boldsymbol{P} \right)
	+\mu \, L_{\text{c}}^2/2 \, \left\lVert \boldsymbol{\nabla P} \right\rVert^2
	$. This formulations is not reconcilable with the relaxed micromorphic model even if we neglect the curvature part.
}.
In that specific case, the simple shear solution for the micro-distortion field $\boldsymbol{P}$ obtains the format
\begin{equation}
\boldsymbol{P} = \left(
\begin{array}{ccc}
0 & P_{12}(x_{2}) & 0 \\
0 & 0 & 0 \\
0 & 0 & 0 \\
\end{array}
\right)
\quad
\mbox{and}
\quad
\boldsymbol{\nabla u} = \left(
\begin{array}{ccc}
0 & u_{1,2}(x_{2}) & 0 \\
0 & 0 & 0 \\
0 & 0 & 0 \\
\end{array}
\right).
\label{eq:non_zero_compo_MM_Mind}
\end{equation}
In general, such a format of the solution is not to be expected.
While it is possible to construct the general simple shear solution to the classical micromorphic model eq.~(\ref{eq:energy_MM_Mind}), for comparison with our other models we consider the energy
\begin{equation}
\begin{split}
W \left(\boldsymbol{\nabla u}, \boldsymbol{P}, \boldsymbol{\nabla P}\right) = & \,
	\dfrac{\mu_{\text{e}} + \mu _{\text{micro}} + \mu_{\text{c}}}{2} \, \left \lVert \boldsymbol{\nabla u} -\boldsymbol{P} \right \rVert ^2
	+ \dfrac{\mu_{\text{e}} + \mu _{\text{micro}} - \mu_{\text{c}}}{2} \, \langle \boldsymbol{\nabla u} - \boldsymbol{P},\left(\boldsymbol{\nabla u} - \boldsymbol{P}\right)^{T} \rangle 
	\\[3mm]
&	+ \dfrac{\lambda_{\text{e}} + \lambda_{\text{micro}}}{2} \, \mbox{tr}^2 \left(\boldsymbol{\nabla u} - \boldsymbol{P}\right) 
	+ \mu _{\text{micro}} \, \left \lVert \mbox{sym} \, \boldsymbol{\nabla u} \right \rVert ^2 
	+ \dfrac{\lambda_{\text{micro}}}{2} \, \mbox{tr}^2 \left(\boldsymbol{\nabla u}\right) 
	\\[3mm]
&	-2\mu _{\text{micro}}  \, \langle \boldsymbol{\nabla u} - \boldsymbol{P},\mbox{sym} \, \boldsymbol{\nabla u} \rangle
	-\lambda_{\text{micro}} \, \mbox{tr} \left(\boldsymbol{\nabla u} - \boldsymbol{P}\right)\mbox{tr} \left(\boldsymbol{\nabla u}\right)
	+\dfrac{\mu \, L_{\text{c}}^2}{2} \, \left \lVert \boldsymbol{\nabla P} \right \rVert ^2
\end{split}
\label{eq:energy_MM_Mind_neff}
\end{equation}
beeing a special case of eq.~(\ref{eq:energy_MM_Mind}), by setting the values of the elastic parameters as follow (see \cite{neff2004material})
\begin{equation}
\begin{array}{cccccccc}
\widehat{\mu} = \mu _{\text{micro}} \, , \quad \widehat{\lambda} = \lambda_{\text{micro}} \, , \quad b_1 = \lambda_{\text{e}} + \lambda_{\text{micro}} \, ,
\quad
b_2 = \mu_{\text{e}} + \mu _{\text{micro}} + \mu_{\text{c}} \, , \quad b_3 = \mu_{\text{e}} + \mu _{\text{micro}} - \mu_{\text{c}} \, , 
\\[3mm]
g_1 = -\lambda_{\text{micro}} \, , \quad g_2 = -2\mu_{\text{micro}} \, ,  \quad a_{10} = \mu \, L_{\text{c}}^2 \, ,
\quad
a_{\{1,2,3,4,5,8,11,13,14,15\}} = 0 \, .
\end{array}
\label{eq:energy_MM_Mind_neff_coeff}
\end{equation}.
We are going to show results for this model in its simplified form eq.~(\ref{eq:energy_MM_Mind_neff}) in Section \ref{sec:MM}.
\section{Micro-void and micro-stretch model}
The expression of the strain energy for the isotropic micro-void continuum with a single curvature parameter (3+1=4 dof's) can be written as:
\begin{equation}
\begin{split}
W \left(\boldsymbol{\nabla u}, \omega ,\mbox{Curl}\,\left(\omega \boldsymbol{\mathbbm{1}}\right) \right) = &
\, \mu _{\text{e}} \left\lVert \mbox{sym} \boldsymbol{\nabla u} - \omega \boldsymbol{\mathbbm{1}} \right\rVert^{2}
+ \dfrac{\lambda_{\text{e}}}{2} \mbox{tr}^2 \left(\boldsymbol{\nabla u} - \omega \boldsymbol{\mathbbm{1}} \right) 
+ \mu_{\text{micro}} \left\lVert \omega \boldsymbol{\mathbbm{1}} \right\rVert^{2}\\[3mm]
&
+ \dfrac{\lambda_{\text{micro}}}{2} \mbox{tr}^2 \left(\omega \boldsymbol{\mathbbm{1}} \right)
+ \dfrac{\mu \,L_{\text{c}}^2}{2} \, \left\lVert \mbox{Curl} \, \left(\omega \boldsymbol{\mathbbm{1}}\right) \right\rVert^2.
\end{split}
\label{eq:energy_MV}
\end{equation}
Here, $\omega : \mathbb{R}^3 \to \mathbb{R}$ describes the additional micro-voids degree of freedom.
On the other hand the expression of the strain energy for the isotropic micro-stretch continuum with a single curvature parameter (3+3+1=7 dof's) can be written as \cite{neff2014unifying}:
\begin{equation}
\begin{split}
W \left(\boldsymbol{\nabla u}, \boldsymbol{A},\omega,\mbox{Curl}\,\left(\boldsymbol{A} - \omega \boldsymbol{\mathbbm{1}}\right)\right) = &
\, \mu _{\text{e}} \left\lVert \mbox{sym} \boldsymbol{\nabla u} - \omega \boldsymbol{\mathbbm{1}} \right\rVert^{2}
+ \dfrac{\lambda_{\text{e}}}{2} \mbox{tr}^2 \left(\boldsymbol{\nabla u} - \omega \boldsymbol{\mathbbm{1}} \right) 
+ \mu _{\text{c}} \left\lVert \mbox{skew} \left(\boldsymbol{\nabla u} - \boldsymbol{A} \right) \right\rVert^{2} \\[3mm]
&
+ \mu_{\text{micro}} \left\lVert \omega \boldsymbol{\mathbbm{1}} \right\rVert^{2}
+ \dfrac{\lambda_{\text{micro}}}{2} \mbox{tr}^2 \left(\omega \boldsymbol{\mathbbm{1}} \right)
+ \dfrac{\mu \,L_{\text{c}}^2}{2} \, \left\lVert \mbox{Curl} \, \left(\boldsymbol{A} + \omega \boldsymbol{\mathbbm{1}}\right)\right\rVert^2,
\end{split}
\label{eq:energy_MS}
\end{equation}
where $\boldsymbol{A} \in \mathfrak{so}(3)$ and $\omega \in \mathbb{R}$.
Both these micro-void and micro-stretch models can be obtained as special cases of the relaxed micromorphic model and because of that, the full solution is not reported in this work.


\section{Simple shear for the isotropic relaxed micromorphic model}
The expression of the strain energy for the isotropic relaxed micromorphic continuum is:
\begin{equation}
\begin{split}
W \left(\boldsymbol{\nabla u}, \boldsymbol{P},\mbox{Curl}\,\boldsymbol{P}\right) = &
  \, \mu _{\text{e}} \left\lVert \mbox{sym} \left(\boldsymbol{\nabla u} - \boldsymbol{P} \right) \right\rVert^{2}
+ \dfrac{\lambda_{\text{e}}}{2} \mbox{tr}^2 \left(\boldsymbol{\nabla u} - \boldsymbol{P} \right) 
+ \mu _{\text{c}} \left\lVert \mbox{skew} \left(\boldsymbol{\nabla u} - \boldsymbol{P} \right) \right\rVert^{2} \\[3mm]
&
+ \mu_{\text{micro}} \left\lVert \mbox{sym}\,\boldsymbol{P} \right\rVert^{2}
+ \dfrac{\lambda_{\text{micro}}}{2} \mbox{tr}^2 \left(\boldsymbol{P} \right)
+ \dfrac{\mu \,L_{\text{c}}^2 }{2} \, \left\lVert \mbox{Curl} \, \boldsymbol{P}\right\rVert^2,
\end{split}
\label{eq:energy}
\end{equation}
while the equilibrium equations without body forces are the following:
\begin{equation}
\begin{array}{rr}
\mbox{Div}\overbrace{\left[2\mu _{\text{e}}\,\mbox{sym} \left(\boldsymbol{\nabla u} - \boldsymbol{P} \right) + \lambda_{\text{e}} \mbox{tr} \left(\boldsymbol{\nabla u} - \boldsymbol{P} \right) \boldsymbol{\mathbbm{1}}
	+ 2\mu _{\text{c}}\,\mbox{skew} \left(\boldsymbol{\nabla u} - \boldsymbol{P} \right)\right]}^{\mathlarger{\widetilde{\boldsymbol{\sigma}}}}
&= \boldsymbol{0},
\\[3mm]
	\widetilde{\sigma}
	- 2 \mu _{\text{micro}}\,\mbox{sym}\,\boldsymbol{P} - \lambda_{\text{micro}} \mbox{tr} \left(\boldsymbol{P}\right) \boldsymbol{\mathbbm{1}}
	- \mu \, L_{\text{c}}^{2}\,\mbox{Curl}\,\mbox{Curl}\,\boldsymbol{P} &= \boldsymbol{0}.
\end{array}
\label{eq:equiMic}
\end{equation}

Note that contrary to the full micromorphic model, the momentum stress tensor $\boldsymbol{m}= \mu \, L_{\text{c}}^2 \, \mbox{Curl} \boldsymbol{P}$ remains of second order due to the Curl operator.
It can be noted that $\boldsymbol{\widetilde{\sigma}}$ can be obtained from eq.~(\ref{eq:equiMic})$_2$ and substituted into eq.~(\ref{eq:equiMic})$_1$. This allows us to obtain the following relation
\begin{equation}
\mbox{Div}\left[ 2 \mu _{\text{micro}}\,\mbox{sym}\,\boldsymbol{P} + \lambda_{\text{micro}} \mbox{tr} \left(\boldsymbol{P}\right) \boldsymbol{\mathbbm{1}} \right] = \boldsymbol{0},
\label{eq:DivS}
\end{equation}
which can be seen as a classical elastic equilibrium equation at the micro-level.

No assumptions will be made on the structure of $\boldsymbol{u}$ and $\boldsymbol{P}$ besides that all the components with index 3 are zero and that the non zero ones depend only on $x_{2}$:
\begin{equation}
\boldsymbol{u} = \left(
\begin{array}{c}
u_{1}(x_{2}) \\
u_{2}(x_{2}) \\
0 \\
\end{array}
\right),
\quad
\boldsymbol{P} = \left(
\begin{array}{ccc}
P_{11}(x_{2}) & P_{12}(x_{2}) & 0 \\
P_{21}(x_{2}) & P_{22}(x_{2}) & 0 \\
0 & 0 & 0 \\
\end{array}
\right).
\label{eq:non_zero_compo}
\end{equation}

The boundary condition for the simple shear are the following:
\begin{equation}
\begin{array}{rrrrrrr}
u_{1}(x_{2} = 0)  = 0 \, , & u_{1}(x_{2} = h)  = \boldsymbol{\gamma} \, h \, , &
u_{2}(x_{2} = 0)  = 0 \, , & u_{2}(x_{2} = h)  = 0 \, , \\[3mm]
P_{21}(x_{2} = 0) = 0 \, , & P_{21}(x_{2} = h) = 0 \, , \;\;\: &
P_{11}(x_{2} = 0) = 0 \, , & P_{11}(x_{2} = h) = 0 \, .\\[3mm]
\end{array}
\label{eq:BC}
\end{equation}
The constraint on the components of $\boldsymbol{P}$ is given by the compatibility condition $\boldsymbol{\nabla u}\cdot \boldsymbol{\tau} = \boldsymbol{P}\cdot \boldsymbol{\tau}$, where $\boldsymbol{\tau}$ is the tangential unit vector on the upper and lower surface.

After substituting the expressions eq.~(\ref{eq:non_zero_compo}) in eq.~(\ref{eq:equiMic}), the non-trivial equilibrium equations reduces to the following two sets
\footnote{
	The second set of equations describe the (plain strain) tension problem, being uncoupled from the shear problem.\\
	The ode eqs.~(\ref{eq:equiSheOther}) are homogeneous as well as the respective boundary conditions eq.~(\ref{eq:BC})$_3$, eq.~(\ref{eq:BC})$_4$, eq.~(\ref{eq:BC})$_7$ and eq.~(\ref{eq:BC})$_8$. 
	Thus, it is clear that the solution is zero (provided that it is unique, which has been demonstrated for the relaxed theory in general).
	The fact that these modes decouple follows from symmetry considerations. The shear problem is point symmetric with respect to the centre whereas the tension problem is double mirror symmetric.
}
\begin{align}
\begin{split}
\left(\mu_{\text{e}}+\mu_{\text{c}}\right) P_{12}'\left(x_2\right) + \left(\mu_{\text{e}}-\mu_{\text{c}}\right) P_{21}'\left(x_2\right) - \left(\mu_{\text{e}}+\mu_{\text{c}}\right) u_1''\left(x_2\right) &= 0\\
\left(\mu_{\text{e}}+\mu _{\text{micro}}+\mu_{\text{c}}\right) P_{12}\left(x_2\right) + \left(\mu_{\text{e}}+\mu _{\text{micro}} -\mu_{\text{c}}\right) P_{21}\left(x_2\right) - \left(\mu_{\text{c}}+\mu_{\text{e}}\right) u_1'\left(x_2\right) &= 0\\
\left(\mu_{\text{e}}+\mu _{\text{micro}}-\mu_{\text{c}}\right) P_{12}\left(x_2\right) + \left(\mu_{\text{e}}+\mu _{\text{micro}}+\mu_{\text{c}}\right) P_{21}\left(x_2\right) - \left(\mu_{\text{e}}-\mu_{\text{c}}\right) u_1'\left(x_2\right) - \mu \, L_{\text{c}}^2 \, P_{21}''\left(x_2\right) &= 0 \, ,
\end{split}
\label{eq:equiShe}
\\[3mm]
\begin{split}
\left(\lambda_{\text{e}}+2 \mu_{\text{e}}\right) \left(P_{22}'\left(x_2\right)-u_{2}''\left(x_2\right)\right)+\lambda_{\text{e}} P_{11}'\left(x_2\right) &= 0 \\
\left(\lambda_{\text{e}}+2 \mu_{\text{e}} + \lambda_{\text{micro}} + 2\mu _{\text{micro}}\right) P_{11}\left(x_2\right) + \left(\lambda_{\text{e}}+\lambda_{\text{micro}}\right)P_{22}\left(x_2\right) + \lambda_{\text{e}} u_{2}'\left(x_2\right) + \mu \, L_{\text{c}}^2 \, P_{11}''\left(x_2\right) &= 0 \\
\left(\lambda_{\text{e}}+2 \mu_{\text{e}} + \lambda_{\text{micro}} + 2\mu _{\text{micro}}\right) P_{22}\left(x_2\right) + \left(\lambda_{\text{e}}+\lambda_{\text{micro}}\right)P_{11}\left(x_2\right) - \left(\lambda_{\text{e}}+2 \mu_{\text{e}}\right) u_{2}'\left(x_2\right) &= 0 \\
\left(\lambda_{\text{e}}+\lambda_{\text{micro}}\right)\left(P_{11}\left(x_2\right) + P_{22}\left(x_2\right)\right) - \lambda_{\text{e}} u_{2}'\left(x_2\right) &= 0 \, .
\end{split}
\label{eq:equiSheOther}
\end{align}
It can be noticed that the first set of equations is uncoupled from the second one: the first depends on $P_{12}(x_2)$, $P_{21}(x_2)$, and $u_{1}(x_2)$, while the second on $P_{11}(x_2)$, $P_{22}(x_2)$, and $u_{2}(x_2)$.

From eq.~(\ref{eq:equiSheOther})$_4$ it is possible to evaluate $u_{2}'(x_2)$ and consequently $u_{2}''(x_2)$
\begin{equation}
	\begin{array}{l}
	u_{2}'(x_2) = 
	\dfrac{\left(\lambda_{\text{e}}+\lambda_{\text{micro}}\right)\left(P_{11}\left(x_2\right) + P_{22}\left(x_2\right)\right)}{\lambda_{\text{e}}},
	\quad
	u_{2}''(x_2) = 
	\dfrac{\left(\lambda_{\text{e}}+\lambda_{\text{micro}}\right)\left(P_{11}'\left(x_2\right) + P_{22}'\left(x_2\right)\right)}{\lambda_{\text{e}}}.
	\end{array}
	\label{eq:u2prime}
\end{equation}
Substituting eq.~(\ref{eq:u2prime}) in eq.~(\ref{eq:equiSheOther})$_3$ gives a linear relation between $P_{11}(x_2)$ and $P_{22}(x_2)$
\begin{equation}
P_{22} (x_2) = 
\dfrac{\mu_{\text{e}}  \left(\lambda_{\text{e}}+\lambda_{\text{micro}}\right)}{\lambda_{\text{e}} \mu _{\text{micro}}-\mu_{\text{e}} \lambda_{\text{micro}}}P_{11}\left(x_2\right).
\label{eq:P11P22}
\end{equation}

After substituting eq.~(\ref{eq:u2prime})$_1$ and the first derivative of eq.~(\ref{eq:P11P22}) in eq.~(\ref{eq:equiSheOther})$_1$ it is possible to obtain the following relation:
\begin{equation}
\dfrac{ \lambda_{\text{micro}} \mu _{\text{micro}} \left(\lambda_{\text{e}} + 2\mu _{\text{e}} \right) + \lambda_{\text{e}} \mu_{\text{e}} \left(\lambda_m + 2\mu_m \right)}{\lambda_{\text{e}} \mu _{\text{micro}}-\mu_{\text{e}} \lambda_{\text{micro}}}P_{11}'\left(x_2\right) = 0 \, ,
\label{eq:P11prime}
\end{equation}
which forces $P_{11}(x_2)$ to be zero given the boundary conditions eq.~(\ref{eq:BC}).
Consequently, $P_{22}(x_2)$ and $u_{2}(x_2)$ have to be zero due to eq.~(\ref{eq:P11P22}), eq.~(\ref{eq:u2prime}) and the boundary conditions on $u_{2}(x_2)$.

From eq.~(\ref{eq:equiShe})$_{2}$ is it possible to derive $u_{1}'(x_{2})$ as a function of $P_{12}(x_{2})$ and $P_{21}(x_{2})$ and thanks to that evaluate $u_{1}''(x_{2})$
\begin{equation}
\begin{split}
u_{1}'(x_{2}) = 
\dfrac{\mu_{\text{c}} \left(P_{12}(x_{2})-P_{21}(x_{2})\right)+\left(\mu_{\text{e}}+\mu _{\text{micro}}\right) \left(P_{12}(x_{2})+P_{21}(x_{2})\right)}{\mu_{\text{c}}+\mu_{\text{e}}},
\\[3mm]
u_{1}''(x_{2}) = 
\dfrac{\mu_{\text{c}} \left(P_{12}'(x_{2})-P_{21}'(x_{2})\right)+\left(\mu_{\text{e}}+\mu _{\text{micro}}\right) \left(P_{12}'(x_{2})+P_{21}'(x_{2})\right)}{\mu_{\text{c}}+\mu_{\text{e}}}.
\end{split}
\label{eq:uprime_usecond}
\end{equation}
Substituting eqs.~(\ref{eq:uprime_usecond}) in eqs.~(\ref{eq:equiShe}) the two remaining equations become
\begin{equation}
\begin{split}
\mu _{\text{micro}} \left(P_{12}'(x_{2})+P_{21}'(x_{2})\right) = 0 \, ,
\\[3mm]
\mu \, L_{\text{c}}^2 \, P_{21}''(x_{2})-\dfrac{2 \mu_{\text{c}} \left(P_{21}(x_{2}) \left(2 \mu_{\text{e}}+\mu _{\text{micro}}\right)+\mu _{\text{micro}} P_{12}(x_{2})\right)}{\mu_{\text{c}}+\mu_{\text{e}}} = 0 \, .
\\[3mm]
\end{split}
\label{eq:equiShe2}
\end{equation}
It can be deduced form eq.~(\ref{eq:equiShe2})$_{1}$ that $P_{12}(x_{2}) = - P_{21}(x_{2}) + C_{0}$ where $C_{0}$ is a constant that will be determined later thanks to the boundary conditions. 
Given the aforementioned relation between the micro-distortions eq.~(\ref{eq:equiShe2}) becomes
\begin{equation}
\mu \, L_{\text{c}}^2 \, P_{21}''(x_{2})-\dfrac{4 \, \mu_{\text{c}} \, \mu_{\text{e}}}{\mu_{\text{c}}+\mu_{\text{e}}} \, P_{21}(x_{2}) + \dfrac{2 \, \mu_{\text{c}} \, \mu _{\text{micro}}}{\mu_{\text{c}}+\mu_{\text{e}}} \, C_{0} = 0 \, ,
\label{eq:equiShe3}
\end{equation}
which is now a simple second order differential equation in $P_{21}(x_{2})$ whose solution is
\begin{equation}
\begin{array}{l}
P_{21}(x_{2}) =
C_{1} \, e^{-\dfrac{2 x_{2} f_1}{L_{\text{c}}}} + C_{2} \, e^{\dfrac{2 x_{2} f_1}{L_{\text{c}}}}-\dfrac{C_{0} \mu _{\text{micro}}}{2 \mu_{\text{e}}} \, ,
\qquad
f_1 \coloneqq \sqrt{\dfrac{\mu _{\text{c}} \mu _{\text{e}}}{\mu \left(\mu _{\text{c}} +\mu _{\text{e}}\right)}} \, .
\end{array}
\label{eq:P12SolDiff}
\end{equation}

The last step before applying the boundary conditions is to calculate $u_{1}(x_{2})$ by integrating eqs.~(\ref{eq:uprime_usecond})$_1$  while substituting the eqs.~(\ref{eq:P12SolDiff}) in it.
The expression of $u_{1}(x_{2})$ is the following:
\begin{equation}
u_{1}(x_{2}) =
\dfrac{\mu_{\text{c}} \, L_{\text{c}}}{f_1 \left(\mu_{\text{c}}+\mu_{\text{e}}\right)}\left(C_{1} \, e^{-\dfrac{2 f_1 x_{2}}{L_{\text{c}}}} - C_{2} \, e^{\dfrac{2 f_1 x_{2}}{L_{\text{c}}}}\right) + \dfrac{\left(\mu_{\text{e}}+\mu _{\text{micro}}\right)}{\mu_{\text{e}}} \, C_{0} \, x_{2} + C_{3}.
\label{eq:uSolDiff}
\end{equation}
After applying the boundary condition eqs.~(\ref{eq:BC}), the values of the integration constants are the following:
\begin{equation}
\begin{array}{l}
C_{0} =
\dfrac{ \cosh \left(\dfrac{f_1 h}{L_{\text{c}}}\right)}{\cosh \left(\dfrac{f_1h}{L_{\text{c}}} \right)-\dfrac{f_2 L_{\text{c}}}{h}\sinh \left(\dfrac{f_1 h}{L_{\text{c}}}\right)}\, \dfrac{\mu_{\text{e}}}{\mu_{\text{e}}+\mu _{\text{micro}}} \, \boldsymbol{\gamma} ,
\\[10mm]
C_{1} =
\dfrac{1}{4}\dfrac{\cosh \left(\dfrac{f_1 h}{L_{\text{c}}}\right) + \sinh \left(\dfrac{f_1 h}{L_{\text{c}}}\right)}{\cosh \left(\dfrac{f_1 h}{L_{\text{c}}}\right)-\dfrac{f_2 L_{\text{c}} }{h}\sinh \left(\dfrac{f_1 h}{L_{\text{c}}}\right)} \, \dfrac{\mu _{\text{micro}}}{\mu_{\text{e}}+\mu _{\text{micro}}} \, \boldsymbol{\gamma} ,
\\[10mm]
C_{2} =
\dfrac{1}{4}\dfrac{\cosh \left(\dfrac{f_1 h}{L_{\text{c}}}\right) - \sinh \left(\dfrac{f_1 h}{L_{\text{c}}}\right)}{\cosh \left(\dfrac{f_1 h}{L_{\text{c}}}\right)-\dfrac{f_2 L_{\text{c}} }{h}\sinh \left(\dfrac{f_1 h}{L_{\text{c}}}\right)} \, \dfrac{\mu _{\text{micro}}}{\mu_{\text{e}}+\mu _{\text{micro}}} \, \boldsymbol{\gamma} ,
\\[10mm]
C_{3} = 
-\dfrac{1}{2}\dfrac{ \sinh \left(\dfrac{f_1 h}{L_{\text{c}}}\right)}{\cosh \left(\dfrac{f_1h}{L_{\text{c}}} \right)-\dfrac{f_2 L_{\text{c}}}{h}\sinh \left(\dfrac{f_1 h}{L_{\text{c}}}\right)} \, f_2 \, L_{\text{c}} \, \boldsymbol{\gamma} .
\end{array}
\label{eq:IntCons}
\end{equation}

\begin{figure}[H]
	\centering
	\includegraphics[width=0.69\textwidth]{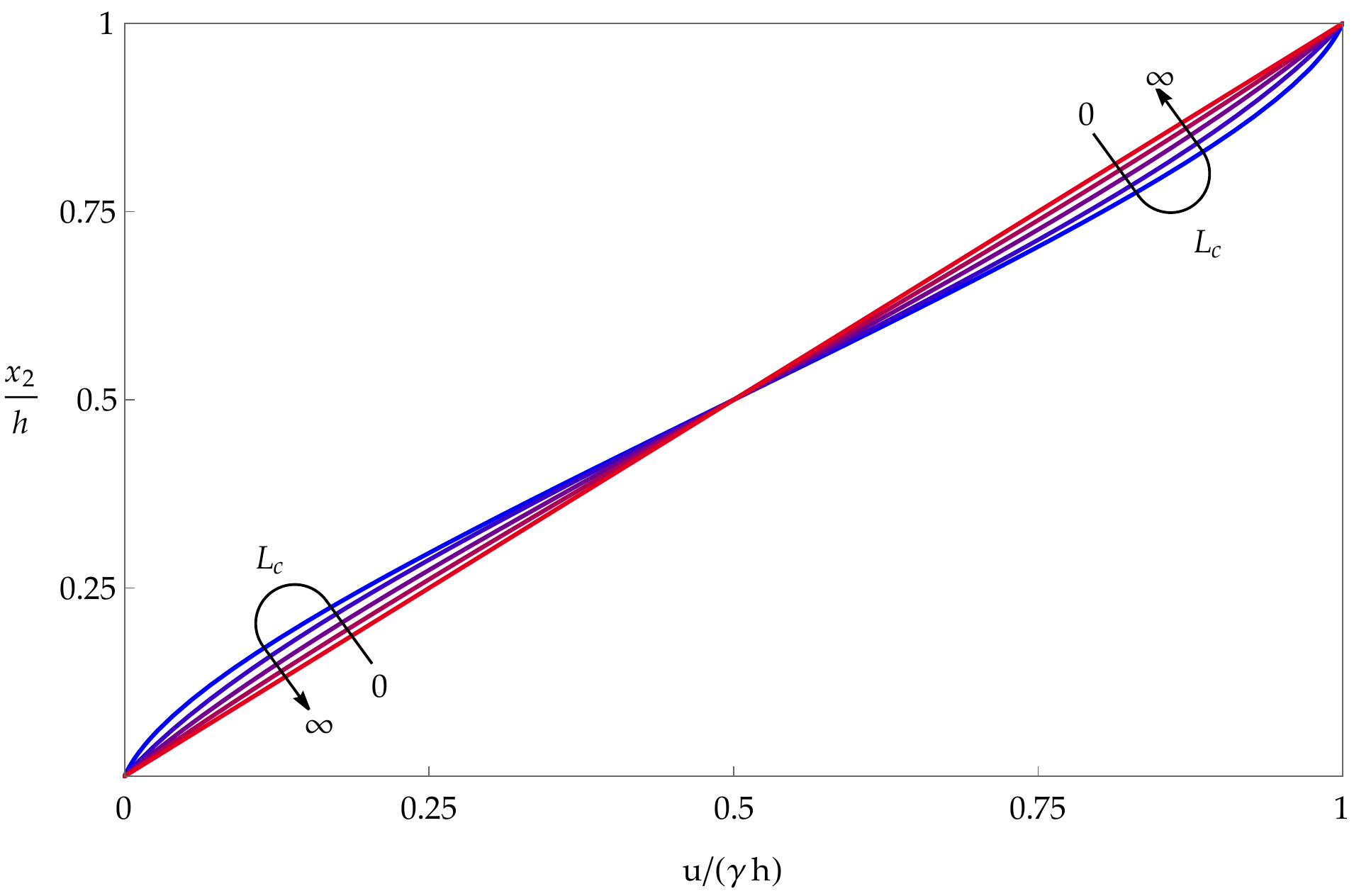}
	\caption{Profile of the dimensionless displacement field for the relaxed micromorphic model for $f_1 = 1.117$, $f_2 = 0.715$ and different values of $L_{\text{c}} = \left\{0.4, 1.25, 2., 3.\overline{3}, 100 \right\}$.
		It is interesting to observe that there appears the maximal inhomogeneity in the displacement not for $L_{\text{c}} \to \infty$ (also equivalent to $h \to 0$) but at an intermediate value.}
	\label{fig:disp_RMM}
\end{figure}

Finally, the expressions of the non-zero displacement field and micro-distortion component are:
\begin{equation}
\begin{array}{rl}
u_{1}(x_{2}) &= \dfrac{\dfrac{f_2 L_{\text{c}} }{h}\sinh \left(\dfrac{f_1 (h-2 x_{2})}{L_{\text{c}}}\right)+\dfrac{2 x_{2} }{h}\cosh \left(\dfrac{f_1 h}{L_{\text{c}}}\right)-\dfrac{f_2 L_{\text{c}}}{h}\sinh \left(\dfrac{f_1 h}{L_{\text{c}}}\right)}{\cosh \left(\dfrac{f_1 h}{L_{\text{c}}}\right)-\dfrac{f_2 L_{\text{c}} }{h}\sinh \left(\dfrac{f_1 h}{L_{\text{c}}}\right)} \dfrac{\boldsymbol{\gamma}  h}{2} ,
\\[1cm]
P_{21}(x_{2}) &=
- \dfrac{\sinh \left(\dfrac{f_1 x_{2}}{L_{\text{c}}}\right) \sinh \left(\dfrac{f_1 (h-x_{2})}{L_{\text{c}}}\right)}{\cosh \left(\dfrac{f_1 h}{L_{\text{c}}}\right)-\dfrac{f_2 L_{\text{c}} }{h} \sinh \left(\dfrac{f_1 h}{L_{\text{c}}}\right)} \, \dfrac{\mu _{\text{micro}} }{\mu_{\text{e}} + \mu _{\text{micro}}} \, \boldsymbol{\gamma} ,
\\[1cm]
P_{12}(x_{2}) &= 
\dfrac{\cosh \left(\dfrac{f_1 h}{L_{\text{c}}}\right) \mu_{\text{e}}  + \sinh \left(\dfrac{f_1 x_2}{L_{\text{c}}}\right) \sinh \left(\dfrac{f_1 (h-x_2)}{L_{\text{c}}}\right) \mu _{\text{micro}} }{ \cosh \left(\dfrac{f_1 h}{L_{\text{c}}}\right)-\dfrac{f_2 L_{\text{c}}}{h}  \sinh \left(\dfrac{f_1 h}{L_{\text{c}}}\right)} \, \dfrac{1}{\mu_{\text{e}}+\mu _{\text{micro}}} \, \boldsymbol{\gamma} ,
\\[1cm]
f_2 &\coloneqq \dfrac{1}{f_1}\dfrac{\mu_{\text{c}} \, \mu _{\text{micro}}}{\left(\mu_{\text{c}}+\mu_{\text{e}}\right) \left(\mu_{\text{e}}+\mu _{\text{micro}}\right)}.
\end{array}
\label{eq:solu_BC}
\end{equation}
A plot of the displacement field while varying $L_{\text{c}}$ is shown in Fig.~\ref{fig:disp_RMM}.

\begin{figure}[H]
	\centering
	\includegraphics[width=0.75\textwidth]{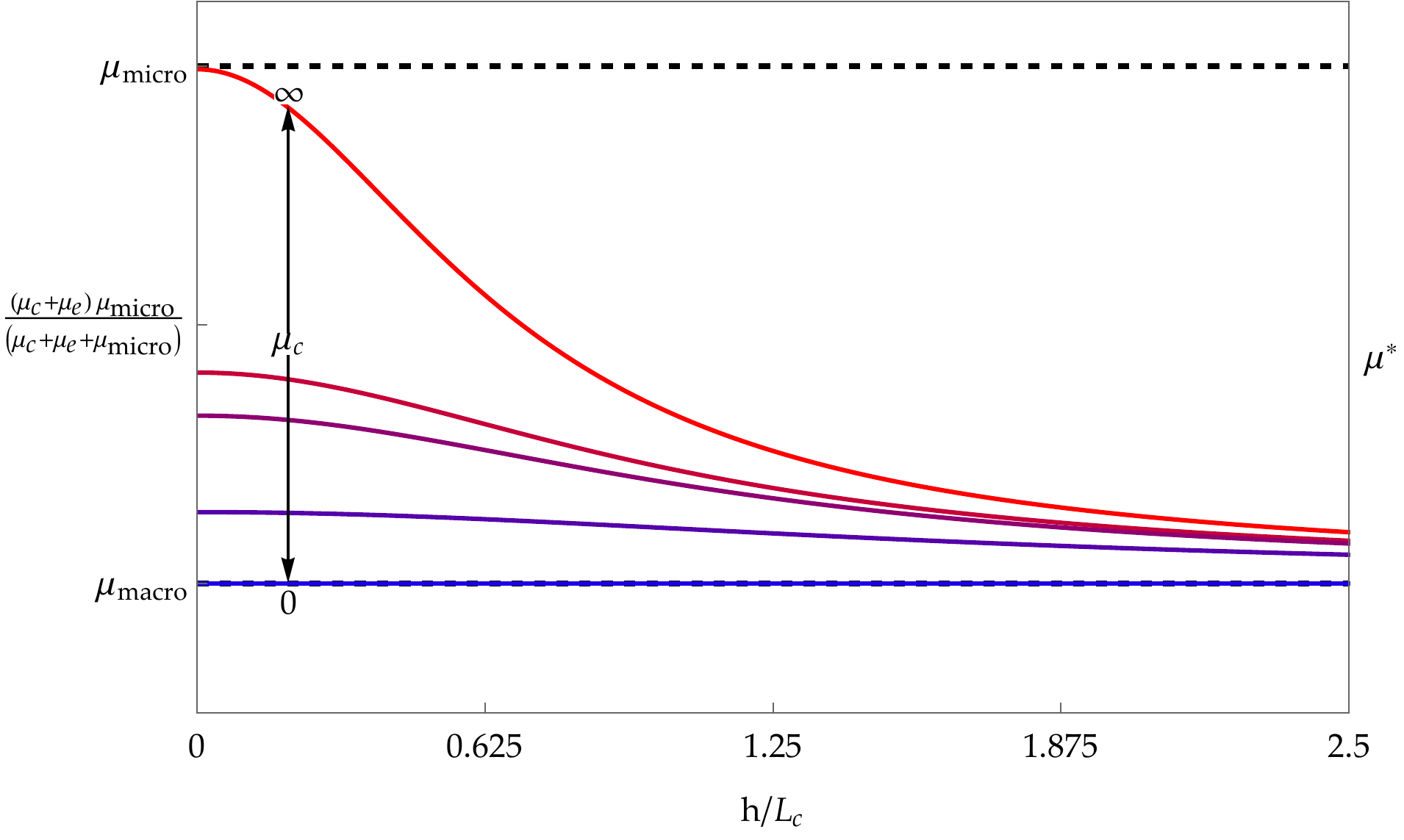}
	\caption{The apparent shear stiffness $\mu^{*}$ governed by $\mu _{\text{macro}}$ (for $L_{\text{c}} = 0$) and $\frac{\left(\mu_{\text{c}} + \mu_{\text{e}}\right) \mu _{\text{micro}}}{\mu_{\text{c}} + \mu_{\text{e}} + \mu _{\text{micro}}}$ (for $L_{\text{c}} \to \infty$).
	In all possible limit cases (except $\mu _{\text{micro}} \to \infty$) the relaxed micromorphic model has bounded shear stiffness.
	The maximum possible shear stiffness is given by $\mu _{\text{micro}}$. The values of the elastic parameters that have been used are: $\mu = 1$, $\mu _{\text{e}} = 1.25$, $\mu _{\text{micro}} = 5$, and $\mu _{\text{c}} = \left\{1000,4.3,3,1,0.00001\right\}$.} 
	\label{fig:stiff_RMM}
\end{figure}

It is worth to be highlighted that $\mbox{sym}\,\boldsymbol{P}$ results to be constant, which satisfies trivially eq.~(\ref{eq:DivS}).
The following relation is a measure of the apparent shear stiffness (see Fig.~\ref{fig:stiff_RMM})
\begin{equation}
\mu^{*} = \dfrac{\widetilde{\sigma}_{12}}{\boldsymbol{\gamma}} = 
\dfrac{1}{1-\dfrac{f_2 L_{\text{c}} }{h}\tanh \left(\dfrac{f_1 h}{L_{\text{c}}}\right)} \dfrac{\mu _{\text{e}} \, \mu _{\text{micro}}}{\mu_ e + \mu _{\text{micro}}}.
\label{eq:stiff_RMM}
\end{equation}

\section{Simple shear for the isotropic Cosserat continuum}
\label{sec:Cos}
The expression of the strain energy for the Cosserat continuum can be written as:
\begin{equation}
\begin{split}
W \left(\boldsymbol{\nabla u}, \boldsymbol{A},\mbox{Curl}\,\boldsymbol{A}\right) = &
  \, \mu _{\text{e}} \left\lVert \mbox{sym} \, \boldsymbol{\nabla u} \right\rVert^{2}
+ \dfrac{\lambda_{\text{e}}}{2} \mbox{tr}^2 \left(\boldsymbol{\nabla u} \right) 
+ \mu _{\text{c}} \left\lVert \mbox{skew} \left(\boldsymbol{\nabla u} - \boldsymbol{A} \right) \right\rVert^{2} 
+ \dfrac{\mu \, L_{\text{c}}^2}{2} \, \left \lVert \mbox{Curl} \, \boldsymbol{A}\right \rVert^2,
\end{split}
\label{eq:energy_Cos}
\end{equation}
where $\boldsymbol{A} \in \mathfrak{so}(3)$.
The equilibrium equations without body forces are the following:
\begin{equation}
\begin{array}{rr}
\mbox{Div}\overbrace{\left[2\mu _{\text{e}}\,\mbox{sym} \, \boldsymbol{\nabla u} + \lambda_{\text{e}} \mbox{tr} \left(\boldsymbol{\nabla u} \right) \boldsymbol{\mathbbm{1}}
	+ 2\mu _{\text{c}}\,\mbox{skew} \left(\boldsymbol{\nabla u} - \boldsymbol{A}\right) \right]}^{\mathlarger{\widetilde{\boldsymbol{\sigma}}}}
&= \boldsymbol{0},
\\[3mm]
2\mu _{\text{c}}\,\mbox{skew} \left(\boldsymbol{\nabla u} - \boldsymbol{A}\right)
- \mu \, L_{\text{c}}^{2} \, \mbox{skew} \, \mbox{Curl}\,\mbox{Curl}\,\boldsymbol{A} &= \boldsymbol{0} \,.
\end{array}
\label{eq:equiMic_Cos}
\end{equation}
This model is a special limit case of the relaxed micromorphic model  for $\mu _{\text{micro}} \to \infty$ .

No assumptions are made on the structure of $\boldsymbol{u}$ and $\boldsymbol{A}$ besides that all the components with index 3 are zero and that the non zero ones depend only on $x_{2}$:
\begin{equation}
\boldsymbol{u} = \left(
\begin{array}{c}
u_{1}(x_{2}) \\
u_{2}(x_{2}) \\
0 \\
\end{array}
\right),
\quad
\boldsymbol{A} = \left(
\begin{array}{cccc}
0 & A_{21}(x_{2}) & 0 \\
- A_{21}(x_{2}) &0 & 0 \\
0 & 0 & 0 \\
\end{array}
\right).
\label{eq:non_zero_compo_Cos}
\end{equation}

The boundary condition for the simple shear are the following:
\begin{equation}
\begin{array}{rrrrrr}
u_{1}(x_{2} = 0) = 0 \, , & u_{1}(x_{2} = h) = \boldsymbol{\gamma} \, h \, , &
u_{2}(x_{2} = 0) = 0 \, , & u_{2}(x_{2} = h) = 0 \, ,\\[3mm]
& A_{21}(x_{2} = 0) = 0 \, , \;\;\: & A_{21}(x_{2} = h) = 0 \,. &
\end{array}
\label{eq:BC_Cos}
\end{equation}
The constraint on the components of $\boldsymbol{A}$ are given by the compatibility condition $\boldsymbol{\nabla u}\cdot \boldsymbol{\tau} = \boldsymbol{A}\cdot \boldsymbol{\tau}$, where $\boldsymbol{\tau}$ is the tangential unit vector on the upper and lower surface.
The resulting combination constrains $\boldsymbol{A}$ to be zero at the upper and lower surface.

After substituting the expressions eq.~(\ref{eq:non_zero_compo_Cos}) in eq.~(\ref{eq:equiMic_Cos}), the non-trivial equilibrium equation reduces to the following two sets
\begin{align}
\left(\mu_{\text{c}} + \mu_{\text{e}}\right)u_{1}''(x_{2}) - 2 \, \mu_{\text{c}} \, A_{21}'(x_{2})   = 0 \, ,
\quad
\dfrac{1}{2} \, \mu \, L_{\text{c}}^2 \, A_{21}''(x_{2}) - 2 \, \mu_{\text{c}} \, A_{21}(x_{2}) + \mu_{\text{c}} \, u_{1}'(x_{2}) & = 0 \, ,
\label{eq:equiShe_Cos}
\\[3mm]
\left(\lambda_{\text{e}}+2 \mu_{\text{e}}\right) u_{2}''\left(x_2\right) &= 0 \footnotemark \, .
\label{eq:equiSheOther_Cos}
\end{align}
\footnotetext{It is noted that for the Cosserat model, the possibility of higher-order effects for the (plane strain) tension problem, which is regulated by eq.~(\ref{eq:equiSheOther_Cos}), is prevented by the fact that this equation does not depend on $L_{\text{c}}$.}
It can be noticed that the first set of equations is uncoupled form the second one: the first depends on $A_{21}(x_2)$ and $u_{1}(x_2)$, while the second only on $u_{2}(x_2)$.

The eq.~(\ref{eq:equiSheOther_Cos}) trivially requires $u_{2}(x_2)$ to be linear, but, due to the boundary condition eq.~(\ref{eq:BC_Cos}), $u_{2}(x_2)$ has to be identically zero.

From eq.~(\ref{eq:equiShe_Cos})$_{1}$ it is possible to derive $u_{1}''(x_{2})$ as a function of $A_{21}'(x_{2})$ and thanks to that evaluate $u_{1}'(x_{2})$
\begin{equation}
u_{1}''(x_{2}) = 
\dfrac{2 \, \mu _{\text{c}}}{\mu_{\text{c}} + \mu_{\text{e}}} A_{21}',
\qquad
u_{1}'(x_{2})  = 
\dfrac{2 \, \mu _{\text{c}}}{\mu_{\text{c}} + \mu_{\text{e}}} A_{21} + C_{0}.
\label{eq:uprime_usecond_Cos}
\end{equation}
where $C_{0}$ is an integration constant which will be determinate with the boundary conditions.

Substituting eqs.~(\ref{eq:uprime_usecond_Cos}) in eqs.~(\ref{eq:equiShe_Cos}), the remaining second order differential equation in $A_{21}(x_{2})$ results to be:
\begin{equation}
\dfrac{1}{2} \, \mu \, L_{\text{c}}^2 A_{21}''(x_{2})  - 2 \, \dfrac{\mu_{\text{c}} \, \mu_{\text{e}}}{\mu_{\text{c}} + \mu_{\text{e}}} A_{21}(x_{2}) + \mu_{\text{c}} \, C_{0} = 0 \, ,
\label{eq:equiShe3_Cos}
\end{equation}
whose solution is 
\begin{equation}
\begin{array}{l}
A_{21}(x_{2}) =
C_{1} \, e^{-\dfrac{2 x_{2} f_1}{L_{\text{c}}}} + C_{2} \, e^{\dfrac{2 x_{2} f_1}{L_{\text{c}}}}-\dfrac{C_{0} \left(\mu_{\text{c}} + \mu_{\text{e}}\right)}{2 \mu_{\text{e}}} \, ,
\qquad
f_1 \coloneqq \sqrt{\dfrac{\mu _{\text{c}} \mu _{\text{e}}}{\mu \left(\mu _{\text{c}} +\mu _{\text{e}}\right)}} \, .
\end{array}
\label{eq:A12SolDiff_Cos}
\end{equation}

Finally, after applying the boundary conditions eq.~(\ref{eq:BC_Cos}) and evaluating the expression of $u_{1}(x_2)$ from eq.~(\ref{eq:uprime_usecond_Cos}), the non-zero displacement field and micro-distortion result to be (see also \cite{neff2009simple,hutter2019micro}):
\begin{equation}
\begin{split}
u_{1} \left(x_2\right) &= \dfrac{\dfrac{\widetilde{f}_2 L_{\text{c}} }{h}\sinh \left(\dfrac{f_1 (h-2 x_{2})}{L_{\text{c}}}\right)+\dfrac{2 x_{2} }{h}\cosh \left(\dfrac{f_1 h}{L_{\text{c}}}\right)-\dfrac{\widetilde{f}_2 L_{\text{c}}}{h}\sinh \left(\dfrac{f_1 h}{L_{\text{c}}}\right)}{\cosh \left(\dfrac{f_1 h}{L_{\text{c}}}\right)-\dfrac{\widetilde{f}_2 L_{\text{c}} }{h}\sinh \left(\dfrac{f_1 h}{L_{\text{c}}}\right)} \dfrac{\boldsymbol{\gamma}  h}{2} ,
\\[3mm]
A_{21} \left(x_2\right) &=
- \dfrac{\sinh \left(\dfrac{f_1 x_{2}}{L_{\text{c}}}\right) \sinh \left(\dfrac{f_1 (h-x_{2})}{L_{\text{c}}}\right)}{\cosh \left(\dfrac{f_1 h}{L_{\text{c}}}\right)-\dfrac{\widetilde{f}_2 L_{\text{c}} }{h} \sinh \left(\dfrac{f_1 h}{L_{\text{c}}}\right)} \, 
\boldsymbol{\gamma},
\qquad
\widetilde{f}_2 \coloneqq \dfrac{1}{f_1}\dfrac{\mu _{\text{c}}}{\mu _{\text{e}} + \mu _{\text{c}}}.
\end{split}
\label{eq:solu_BC_Cos}
\end{equation}

\vspace{3mm}

\begin{figure}[H]
	\centering
	\includegraphics[width=0.75\textwidth]{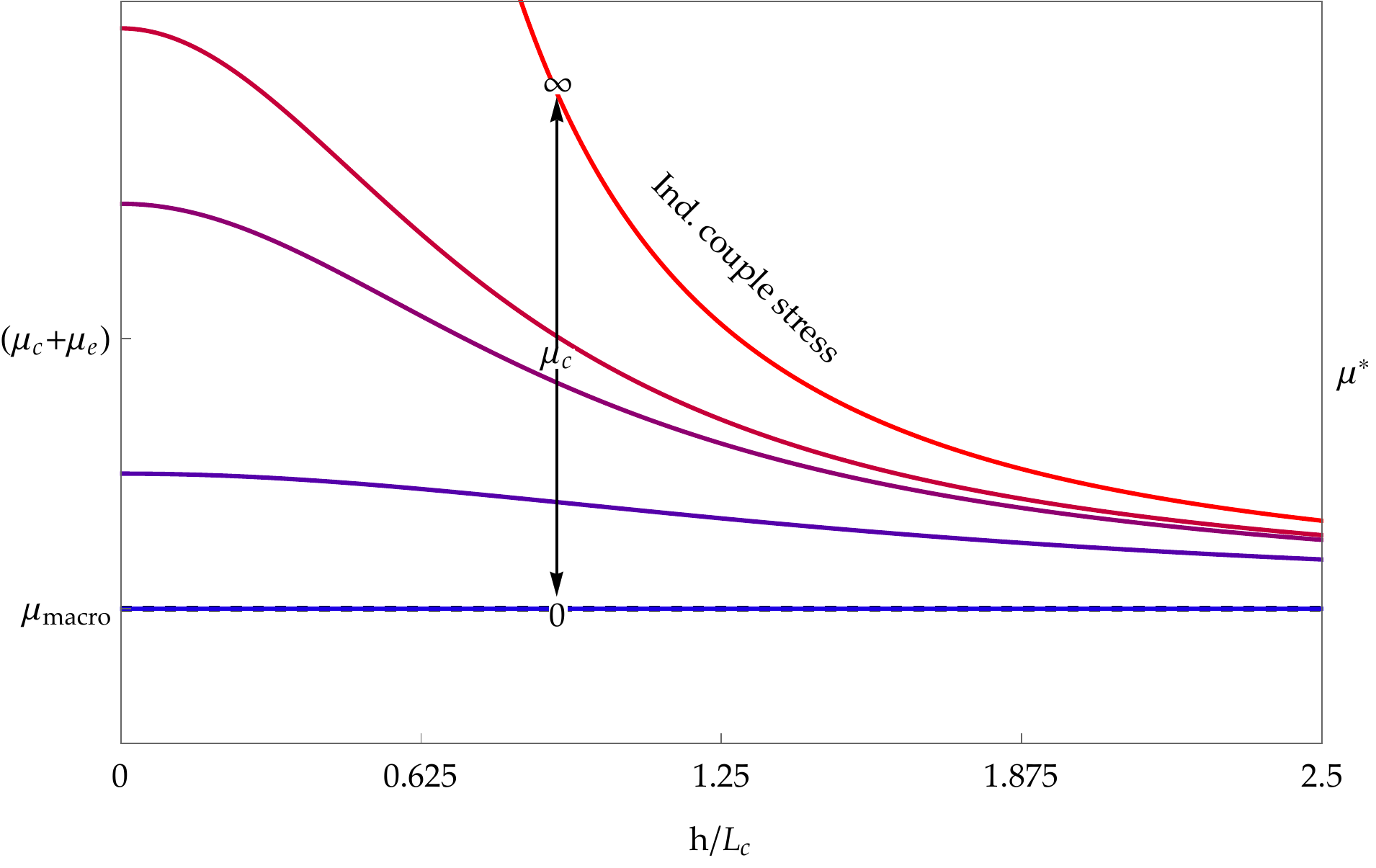}
	\caption{The Cosserat model and its shear stiffness response $\mu^{*}$ as a function of $L_{\text{c}}$.
			For $0 < \mu _{\text{c}} < \infty$ we observe bounded shear stiffness, for all values of the characteristic size $L_{\text{c}}$.
			For $\mu _{\text{c}} \to \infty$ the solution of the indeterminate couple stress model is retrieved for which the shear stiffness is singular for $L_{\text{c}} \to \infty$ ($h \to 0$) due to the applied boundary conditions.
			Also, $\mu_{\text{e}} \to \infty$ generate a singularity for $L_{\text{c}} \to \infty$. The values of the elastic parameters that have been used are: $\mu = 1$, $\mu _{\text{e}} = 1.25$, and $\mu _{\text{c}} = \left\{1000,4.3,3,1,0.00001\right\}$.} 
	\label{fig:stiff_Cos_Couple}
\end{figure}

The following relation is a measure of the higher order stiffness (see Fig.\ref{fig:stiff_Cos_Couple})
\begin{equation}
\mu^{*} = 
\dfrac{1}{1-\dfrac{\widetilde{f}_2 L_{\text{c}} }{h}\tanh \left(\dfrac{f_1 h}{L_{\text{c}}}\right)} \, \mu _{\text{e}}.
\label{eq:stiff_Cos}
\end{equation}

\section{Simple shear for the isotropic indeterminate couple stress continuum}
\label{sec:IndCoupStress}
From the Cosserat model we obtain the indeterminate couple stress model by constraining $\boldsymbol{A} = \mbox{skew} \boldsymbol{\nabla u}$.
The expression of the strain energy for the indeterminate couple stress continuum is:
\begin{equation}
W \left(\boldsymbol{\nabla u}, \mbox{Curl skew} \boldsymbol{\nabla u}\right) = 
\mu_{\text{e}}\left\lVert \mbox{sym} \boldsymbol{\nabla u} \right\rVert^{2} + 
\dfrac{\lambda_{\text{e}}}{2} \mbox{tr}^2\left(\boldsymbol{\nabla u}\right)
+ \dfrac{\mu \,L_{\text{c}}^2}{2} \, \left \lVert \mbox{Curl}\,\mbox{skew}\,\boldsymbol{\nabla u} \right \rVert^2,
\label{eq:energy_Ind}
\end{equation}
while the equilibrium equations without body forces are the following:
\begin{equation}
\mbox{Div}\left[2\mu_{\text{e}} \mbox{sym} \boldsymbol{\nabla u} + 
\lambda_{\text{e}} \mbox{tr} \left(\boldsymbol{\nabla u}\right) \boldsymbol{\mathbbm{1}} 
+ \mu \, L_{\text{c}}^2\,\mbox{skew}\,\mbox{Curl}\,\mbox{Curl}\,\mbox{skew} \boldsymbol{\nabla u}\right] 
= \boldsymbol{0}.
\label{eq:equiMic_Ind}
\end{equation}
This model is a special limit case of the Cosserat model for Cosserat couple modulus  $\mu_{\text{c}} \to \infty$.

No assumptions are made on the structure of $\boldsymbol{u} = \big(u_{1}(x_{2}) ,u_{2}(x_{2}) ,0 \big)^T.$ besides that the third component is zero and that the non zero ones depend only on $x_{2}$.
The boundary condition for the simple shear are the following:
\begin{equation}
\begin{array}{rrrrrrrrrrr}
u_{1}(x_{2} = 0) = 0 \, , &  u_{1}(x_{2} = h) = \boldsymbol{\gamma} \, h \, , &
u_{2}(x_{2} = 0) = 0 \, , &  u_{2}(x_{2} = h) = 0 \, , \\[3mm]
&u_{1}'(x_{2} = 0)  =0 \, , \;\;\: & u_{1}'(x_{2}= h) = 0\, . &
\end{array}
\label{eq:BC_Ind}
\end{equation}
The constraint on the components of $\boldsymbol{\nabla u}$ are given by the compatibility condition $\mbox{skew} \, \boldsymbol{\nabla u}\cdot \boldsymbol{\tau} = \boldsymbol{0}$, where $\boldsymbol{\tau}$ is the tangential unit vector on the upper and lower surface.

After substituting the displacement field in eq.~(\ref{eq:equiMic_Ind}), the non-trivial equilibrium equations reduces to the following two sets
\begin{equation}
\mu_{\text{e}} \, u_{1}''(x_{2}) - \dfrac{1}{4} \, \mu \, L_{\text{c}}^{2} \, u_{1}^{(4)}(x_{2})   = 0 \, ,
\quad \mbox{and} \quad
\left(\lambda_{\text{e}}+2 \mu_{\text{e}}\right) u_{2}''\left(x_2\right) = 0 \, . \footnotemark
\label{eq:equiShe_Other_Ind}
\end{equation}
\footnotetext{It is noted that for the indeterminate couple stress model, the possibility of higher-order effects for the (plane strain) tension problem, which is regulated by eq.~(\ref{eq:equiShe_Other_Ind})$_2$, is prevented by the fact that this equation does not depend on $L_{\text{c}}$.}
It can be noticed that the first equation is uncoupled from the second one: the first depends on $u_{1}(x_2)$, while the second only on $u_{2}(x_2)$.
Eq.~(\ref{eq:equiShe_Other_Ind})$_{2}$ trivially requires $u_{2}(x_2)$ to be linear, but, due to the boundary condition eq.~(\ref{eq:BC_Ind}), $u_{2}(x_2)$ has to be identically zero.

The eq.~(\ref{eq:equiShe_Other_Ind})$_{1}$ is a fourth order differential equation in $u_{1}(x_{2})$ whose solution is 
\begin{equation}
\begin{array}{ll}
u_{1}(x_{2}) =
\dfrac{L_{\text{c}}^{2}}{4 \, \overline{f}_{1}}\left(C_{1} \, e^{-\dfrac{2 x_{2} \overline{f}_1}{L_{\text{c}}}} + C_{2} \, e^{-\dfrac{2 x_{2} \overline{f}_1}{L_{\text{c}}}}\right) + C_{3} \, x_{2} + C_{4} \, ,
\qquad
\overline{f}_1 \coloneqq \sqrt{\dfrac{\mu _{\text{e}}}{\mu}} \,.
\end{array}
\label{eq:P12SolDiff_Ind}
\end{equation}

Finally, after applying the boundary conditions eq.~(\ref{eq:BC_Ind}), the expression of the non-zero component of the displacement field results to be:
\begin{equation}
u_{1} \left(x_2\right) = \dfrac{\dfrac{ L_{\text{c}} }{\overline{f}_{1} \, h}\sinh \left(\dfrac{\overline{f}_1 (h-2 x_{2})}{L_{\text{c}}}\right)+\dfrac{2 x_{2} }{h}\cosh \left(\dfrac{\overline{f}_1 h}{L_{\text{c}}}\right)-\dfrac{ L_{\text{c}}}{\overline{f}_{1} \, h}\sinh \left(\dfrac{\overline{f}_1 h}{L_{\text{c}}}\right)}{\cosh \left(\dfrac{\overline{f}_1 h}{L_{\text{c}}}\right)-\dfrac{ L_{\text{c}} }{\overline{f}_{1} \, h}\sinh \left(\dfrac{\overline{f}_1 h}{L_{\text{c}}}\right)} \dfrac{\boldsymbol{\gamma}  h}{2}.
\label{eq:solu_BC_Ind}
\end{equation}

The following relation is a measure of the higher order stiffness and it can be seen in Fig.~\ref{fig:stiff_Cos_Couple}
\begin{equation}
\mu^{*} = 
\dfrac{1}{1-\dfrac{L_{\text{c}} }{\overline{f}_{1} \, h}\tanh \left(\dfrac{\overline{f}_1 h}{L_{\text{c}}}\right)} \, \mu _{\text{e}} \, .
\label{eq:stiff_Ind}
\end{equation}
\subsection{Simple shear for the isotropic symmetric couple stress continuum}
The expression of the strain energy for the isotropic symmetric couple stress continuum is:
\begin{equation}
W \left(\boldsymbol{\nabla u}, \mbox{Curl sym} \boldsymbol{\nabla u}\right) = 
\mu_{\text{e}}\left\lVert \mbox{sym} \boldsymbol{\nabla u} \right\rVert^{2} + 
\dfrac{\lambda_{\text{e}}}{2} \mbox{tr}^2\left(\boldsymbol{\nabla u}\right)
+ \dfrac{\mu \,L_{\text{c}}^2}{2} \, \left \lVert \mbox{Curl}\,\mbox{sym}\,\boldsymbol{\nabla u} \right \rVert^2,
\label{eq:energy_Ind_sym}
\end{equation}
while the equilibrium equations without body forces are the following:
\begin{equation}
\mbox{Div}\left[2\mu_{\text{e}} \mbox{sym} \boldsymbol{\nabla u} + 
\lambda_{\text{e}} \mbox{tr} \left(\boldsymbol{\nabla u}\right) \boldsymbol{\mathbbm{1}} 
+ \mu \, L_{\text{c}}^2\,\mbox{sym}\,\mbox{Curl}\,\mbox{Curl}\,\mbox{sym} \boldsymbol{\nabla u}\right] 
= \boldsymbol{0}.
\label{eq:equiMic_Ind_sym}
\end{equation}

As it is shown in Appendix \ref{Sec:appendixB}, $ \mbox{Curl}\,\mbox{sym}\,\boldsymbol{\nabla u} = - \mbox{Curl}\,\mbox{skew}\,\boldsymbol{\nabla u} $, which implies that the energy for this model, eq.~(\ref{eq:energy_Ind_sym}), is the same as the indeterminate couple stress continuum energy eq.~(\ref{eq:energy_Ind}).

Observe as well that eq.~(\ref{eq:energy_Ind_sym}) can be seen as a degenerate case of Mindlin's strain gradient continuum eq.~(\ref{eq:energy_SGM}) (Form I of \cite{mindlin1964micro}).

Furthermore, these two models share the same equilibrium equations (eq.~(\ref{eq:equiMic_Ind}) and eq.~(\ref{eq:equiMic_Ind_sym})) since it is possible to prove that $\mbox{Div}\left[\mbox{sym}\,\mbox{Curl}\,\mbox{Curl}\,\mbox{sym} \boldsymbol{\nabla u}\right] = \mbox{Div}\left[\mbox{skew}\,\mbox{Curl}\,\mbox{Curl}\,\mbox{skew} \boldsymbol{\nabla u}\right]$.

The following expressions show how, in general, the boundary conditions of the symmetric couple stress continuum model ($\mbox{sym} \, \boldsymbol{\nabla u}\cdot \boldsymbol{\tau} = \boldsymbol{0}$) differ from the ones for the indeterminate couple stress model ($\mbox{skew} \, \boldsymbol{\nabla u}\cdot \boldsymbol{\tau} = \boldsymbol{0}$):

\begin{equation}
\mbox{sym} \, \boldsymbol{\nabla u}\cdot \boldsymbol{\tau} = \dfrac{1}{2}
\left(
\begin{array}{c}
2 \, u_{1,1} \\
u_{1,2}+u_{2,1} \\
u_{1,3}+u_{3,1} \\
\end{array}
\right) \, ,
\quad
\mbox{skew} \, \boldsymbol{\nabla u}\cdot \boldsymbol{\tau} = \dfrac{1}{2}
\left(
\begin{array}{c}
0 \\
u_{2,1}-u_{1,2} \\
u_{3,1}-u_{1,3} \\
\end{array}
\right) \, .
\label{eq:BC_Ind_vs_Ind_sym}
\end{equation}

However, for the simple shear problem, since $u_{2} = 0$ and the displacement field does not depend on the coordinate $x_1$, eq.~(\ref{eq:BC_Ind_vs_Ind_sym})$_1$ and eq.~(\ref{eq:BC_Ind_vs_Ind_sym})$_2$ end up to be the same:

\begin{equation}
  \mbox{sym} \, \boldsymbol{\nabla u}\cdot \boldsymbol{\tau} =
- \mbox{sym} \, \boldsymbol{\nabla u}\cdot \boldsymbol{\tau} =
\left(
0 ,
\dfrac{u_{1,2}}{2}  ,
0 
\right)^T \, .
\label{eq:BC_Ind_vs_Ind_sym_shear}
\end{equation}
making this model completely equivalent to the indeterminate couple stress model for the simple shear problem.
\subsection{Simple shear for the isotropic modified couple stress continuum and for the isotropic ``pseudo-consistent'' couple stress continuum}
The expression of the strain energy for the isotropic modified couple stress continuum is:
\begin{equation}
W \left(\boldsymbol{\nabla u}, \mbox{Curl skew} \boldsymbol{\nabla u}\right) = 
\mu_{\text{e}}\left\lVert \mbox{sym} \boldsymbol{\nabla u} \right\rVert^{2} + 
\dfrac{\lambda_{\text{e}}}{2} \mbox{tr}^2\left(\boldsymbol{\nabla u}\right)
+ \dfrac{\mu \, \overline{L_{\text{c}}}^2}{2} \, \left \lVert \mbox{sym}\,\mbox{Curl}\,\mbox{skew}\,\boldsymbol{\nabla u} \right \rVert^2 \, .
\label{eq:energy_Ind_M}
\end{equation}
This energy is equivalent to the one of the isotropic indeterminate couple stress continuum beside a constant multiplying the curvature term (expressed via the different notation for length-scale parameter $\overline{L_{\text{c}}}$) and this can be seen thanks to the calculation reported in Appendix \ref{Sec:appendixC}.

The expression of the strain energy for the ``pseudo-consistent'' couple stress continuum is \cite{hadjesfandiari2011couple,neff2016some}:
\begin{equation}
W \left(\boldsymbol{\nabla u}, \mbox{Curl skew} \boldsymbol{\nabla u}\right) = 
\mu_{\text{e}}\left\lVert \mbox{sym} \boldsymbol{\nabla u} \right\rVert^{2} + 
\dfrac{\lambda_{\text{e}}}{2} \mbox{tr}^2\left(\boldsymbol{\nabla u}\right)
+ \dfrac{\mu \, \overline{L_{\text{c}}}^2}{2} \, \left \lVert \mbox{skew}\,\mbox{Curl}\,\mbox{skew}\,\boldsymbol{\nabla u} \right \rVert^2.
\label{eq:energy_Ind_PS}
\end{equation}

The authors of \cite{neff2016some} have argued that the curvature term in the couple stress model should only depend on $\left \lVert \mbox{skew}\,\mbox{Curl}\,\mbox{skew}\,\boldsymbol{\nabla u} \right \rVert$.
In Appendix \ref{Sec:appendixC}, it is possible to see that for simple shear, the solution coincides with the modified couple stress model and the indeterminate couple stress model.
Differences are expected to appear for other boundary value problems like bending.

\section{Simple shear for the classical isotropic micromorphic continuum without mixed terms}
\label{sec:MM}
The expression of the strain energy for the classical isotropic micromorphic continuum without mixed terms (like $\langle\mbox{sym} \boldsymbol{P}, \mbox{sym}\left(\boldsymbol{\nabla u} -\boldsymbol{P}\right)\rangle$, etc.) can be written as:
\begin{equation}
\begin{split}
W \left(\boldsymbol{\nabla u}, \boldsymbol{P}, \boldsymbol{\nabla P}\right) = &
\, \mu _{\text{e}} \left\lVert \mbox{sym} \left(\boldsymbol{\nabla u} - \boldsymbol{P} \right) \right\rVert^{2}
+ \dfrac{\lambda_{\text{e}}}{2} \mbox{tr}^2 \left(\boldsymbol{\nabla u} - \boldsymbol{P} \right)
+ \mu _{\text{c}} \left\lVert \mbox{skew} \left(\boldsymbol{\nabla u} - \boldsymbol{P} \right) \right\rVert^{2} \\[3mm]
&
+ \mu_{\text{micro}} \left\lVert \mbox{sym}\,\boldsymbol{P} \right\rVert^{2}
+ \dfrac{\lambda_{\text{micro}}}{2} \mbox{tr}^2 \left(\boldsymbol{P} \right) 
+ \dfrac{\mu \, L_{\text{c}}^2}{2} \left\lVert \boldsymbol{\nabla P}\right\rVert^2,
\end{split}
\label{eq:energy_MM}
\end{equation}
while the equilibrium equations without body forces are the following:
\begin{equation}
\begin{array}{rr}
\mbox{Div}\overbrace{\left[2\mu _{\text{e}}\,\mbox{sym} \left(\boldsymbol{\nabla u} - \boldsymbol{P} \right) + \lambda_{\text{e}} \mbox{tr} \left(\boldsymbol{\nabla u} - \boldsymbol{P} \right) \boldsymbol{\mathbbm{1}}
	+ 2\mu _{\text{c}}\,\mbox{skew} \left(\boldsymbol{\nabla u} - \boldsymbol{P} \right)\right]}^{\mathlarger{\widetilde{\boldsymbol{\sigma}}}}
&= \boldsymbol{0} \, ,
\\[3mm]
\widetilde{\sigma}
- 2 \mu _{\text{micro}}\,\mbox{sym}\,\boldsymbol{P} - \lambda_{\text{micro}} \mbox{tr} \left(\boldsymbol{P}\right) \boldsymbol{\mathbbm{1}}
+\mu L_{\text{c}}^{2}\,\mbox{Div} \, \left[\boldsymbol{\nabla P}\right] &= \boldsymbol{0} \, .
\end{array}
\label{eq:equiMic_MM}
\end{equation}
Observe that the momentum stress tensor $\boldsymbol{m} = \mu \, L_{\text{c}}^2 \, \boldsymbol{\nabla} \boldsymbol{P}$ is of third order.
No assumptions are made on the structure of $\boldsymbol{u}$ and $\boldsymbol{P}$ besides that all the components with index 3 are zero and that the non zero ones depend only on $x_{2}$:
\begin{equation}
\boldsymbol{u} = \left(
\begin{array}{c}
u_{1}(x_{2}) \\
u_{2}(x_{2}) \\
0 \\
\end{array}
\right),
\quad
\boldsymbol{P} = \left(
\begin{array}{ccc}
P_{11}(x_{2}) & P_{12}(x_{2}) & 0 \\
P_{21}(x_{2}) & P_{22}(x_{2}) & 0 \\
0 & 0 & 0 \\
\end{array}
\right).
\label{eq:non_zero_compo_MM}
\end{equation}

The boundary condition for the simple shear are the following:
\begin{equation}
\begin{array}{rrrrrrrrrrr}
u_{1}(x_{2} = 0) = 0 \, , & u_{1}(x_{2} = h) = \boldsymbol{\gamma} \, h \, , &
u_{2}(x_{2} = 0) = 0 \, , & u_{2}(x_{2} = h) = 0 \, , \\[3mm]
&\boldsymbol{P}(x_{2} = 0) = 0 \, , \;\;\: & \boldsymbol{P}(x_{2} = h) = 0\, . &
\end{array}
\label{eq:BC_MM}
\end{equation}
It is important to underline that the constraint is applied on all the components of $\boldsymbol{P}$ at the upper and lower surface.
After substituting the expressions eq.~(\ref{eq:non_zero_compo_MM}) in eq.~(\ref{eq:equiMic_MM}), the non-trivial equilibrium equations reduces to the following two sets
\begin{align}
\begin{split}
\left(\mu_{\text{e}}+\mu_{\text{c}}\right) P_{12}'\left(x_2\right) + \left(\mu_{\text{e}}-\mu_{\text{c}}\right) P_{21}'\left(x_2\right) - \left(\mu_{\text{e}}+\mu_{\text{c}}\right) u_1''\left(x_2\right) &= 0\\
\left(\mu_{\text{e}}+\mu _{\text{micro}}+\mu_{\text{c}}\right) P_{12}\left(x_2\right) + \left(\mu_{\text{e}}+\mu _{\text{micro}} -\mu_{\text{c}}\right) P_{21}\left(x_2\right) - \left(\mu_{\text{c}}+\mu_{\text{e}}\right) u_1'\left(x_2\right)
- \mu \, L_{\text{c}}^2 \, P_{12}''\left(x_2\right) &= 0\\
\left(\mu_{\text{e}}+\mu _{\text{micro}}-\mu_{\text{c}}\right) P_{12}\left(x_2\right) + \left(\mu_{\text{e}}+\mu _{\text{micro}}+\mu_{\text{c}}\right) P_{21}\left(x_2\right) - \left(\mu_{\text{e}}-\mu_{\text{c}}\right) u_1'\left(x_2\right) 
- \mu \, L_{\text{c}}^2 \, P_{21}''\left(x_2\right) &= 0 \, ,
\end{split}
\label{eq:equiShe_MM}
\\[3mm]
\begin{split}
\left(\lambda_{\text{e}}+2 \mu_{\text{e}}\right) \left(P_{22}'\left(x_2\right)-u_{2}''\left(x_2\right)\right)+\lambda_{\text{e}} P_{11}'\left(x_2\right) &= 0 \\
\left(\lambda_{\text{e}}+2 \mu_{\text{e}} + \lambda_{\text{micro}} + 2\mu _{\text{micro}}\right) P_{11}\left(x_2\right) + \left(\lambda_{\text{e}}+\lambda_{\text{micro}}\right)P_{22}\left(x_2\right) + \lambda_{\text{e}} u_{2}'\left(x_2\right)
+ \mu \, L_{\text{c}}^2 \, P_{11}''\left(x_2\right) &= 0 \\
\left(\lambda_{\text{e}}+2 \mu_{\text{e}} + \lambda_{\text{micro}} + 2\mu _{\text{micro}}\right) P_{22}\left(x_2\right) + \left(\lambda_{\text{e}}+\lambda_{\text{micro}}\right)P_{11}\left(x_2\right) - \left(\lambda_{\text{e}}+2 \mu_{\text{e}}\right) u_{2}'\left(x_2\right)  \hspace{1.075cm}\\
+ \mu \, L_{\text{c}}^2 \, P_{22}''\left(x_2\right) &= 0 \\
\left(\lambda_{\text{e}}+\lambda_{\text{micro}}\right)\left(P_{11}\left(x_2\right) + P_{22}\left(x_2\right)\right) - \lambda_{\text{e}} u_{2}'\left(x_2\right) &= 0 \, .
\end{split}
\label{eq:equiSheOther_MM}
\end{align}
It can be noticed that the first set of equations is uncoupled form the second one: the first depends on $P_{12}(x_2)$, $P_{21}(x_2)$, and $u_{1}(x_2)$, while the second on $P_{11}(x_2)$, $P_{22}(x_2)$, and $u_{2}(x_2)$.

From the eq.~(\ref{eq:equiSheOther_MM})$_4$ is it possible to evaluate $u_{2}'(x_2)$ and consequently $u_{2}''(x_2)$
\begin{equation}
\begin{array}{l}
u_{2}'(x_2) = 
\dfrac{\left(\lambda_{\text{e}}+\lambda_{\text{micro}}\right)\left(P_{11}\left(x_2\right) + P_{22}\left(x_2\right)\right)}{\lambda_{\text{e}}},
\quad
u_{2}''(x_2) = 
\dfrac{\left(\lambda_{\text{e}}+\lambda_{\text{micro}}\right)\left(P_{11}'\left(x_2\right) + P_{22}'\left(x_2\right)\right)}{\lambda_{\text{e}}}.
\end{array}
\label{eq:u2prime_MM}
\end{equation}
Substituting eq.~(\ref{eq:u2prime_MM}) in eq.~(\ref{eq:equiSheOther_MM})$_2$ gives a second order differential equation in $P_{11}(x_2)$ which, due to the boundary conditions eqs.~(\ref{eq:BC_MM}), requires $P_{11}(x_2)$ to be identically zero.

After putting to zero $P_{11}(x_2)$ and its derivative eq.~(\ref{eq:equiSheOther_MM})$_3$ becomes a second order differential equation in $P_{22}(x_2)$ which, due to the boundary conditions eqs.~(\ref{eq:BC_MM}), requires $P_{22}(x_2)$ to be identically zero.

It is now clear that $u_{2}(x_2)$ has to be zero due to the fact that $P_{11}(x_2)$ and $P_{22}(x_2)$ are zero, eq.~(\ref{eq:u2prime_MM}) and the boundary conditions eqs.~(\ref{eq:BC_MM}).

From eq.~(\ref{eq:equiShe_MM})$_{1}$ it is possible to derive $u_{1}''(x_{2})$ as a function of $P_{12}(x_{2})$ and $P_{21}(x_{2})$ and thanks to that evaluate $u_{1}'(x_{2})$
\begin{equation}
u_{1}''(x_{2}) = 
P_{12}'(x_{2}) + \dfrac{\left(\mu_{\text{e}} - \mu_{\text{c}} \right)}{\mu_{\text{e}} + \mu_{\text{c}}}P_{21}'(x_{2}),
\quad
u_{1}'(x_{2})  = 
P_{12}(x_{2}) + \dfrac{\left(\mu_{\text{e}} - \mu_{\text{c}} \right)}{\mu_{\text{e}} + \mu_{\text{c}}}P_{21}(x_{2}) + C_{0},
\label{eq:uprime_usecond_MM}
\end{equation}
where $C_0$ is an integration constant which will be evaluated thanks to the boundary conditions.

Substituting eqs.~(\ref{eq:uprime_usecond_MM}) in eqs.~(\ref{eq:equiShe_MM}) the two remaining equations become
\begin{equation}
\begin{split}
- C_0 \left(\mu_{\text{c}}+\mu_{\text{e}}\right) + \mu _{\text{micro}} \left(P_{12}(x_{2})+P_{21}(x_{2})\right) - \mu \, L_{\text{c}}^2 \, P_{12}''(x_{2}) &= 0 \, , \\[3mm]
- C_0 \left(\mu_{\text{e}} - \mu_{\text{c}}\right) + \mu _{\text{micro}} \, P_{12}(x_{2}) + \left(\dfrac{4 \mu_{\text{c}} \mu_{\text{e}}}{\mu_{\text{c}}+\mu_{\text{e}}}+\mu _{\text{micro}}\right) P_{21}(x_{2}) - \mu \, L_{\text{c}}^2 \, P_{21}''(x_{2}) &= 0 \, .
\end{split}
\label{eq:equiShe2_MM}
\end{equation}
Form eq.~(\ref{eq:equiShe2_MM})$_{2}$ it is possible to derive $P_{12}(x_{2})$ as a function of $P_{12}(x_{2})$ and its derivative:
\begin{equation}
\begin{split}
P_{12}(x_{2})     &= 
\dfrac{\left(\mu_{\text{e}} - \mu_{\text{c}}\right)}{\mu _{\text{micro}}} \, C_0 
- \left(\dfrac{4 \mu_{\text{c}} \mu_{\text{e}}}{\left(\mu_{\text{c}}+\mu_{\text{e}}\right)\mu _{\text{micro}}} + 1\right) P_{21}(x_{2})
+ \dfrac{\mu \, L_{\text{c}}^2}{\mu _{\text{micro}}} \, P_{21}''(x_{2}),
\\[3mm]
P_{12}'(x_{2})    &=
- \left(\dfrac{4 \mu_{\text{c}} \mu_{\text{e}}}{\left(\mu_{\text{c}}+\mu_{\text{e}}\right)\mu _{\text{micro}}} + 1\right) P_{21}'(x_{2})
+ \dfrac{\mu \, L_{\text{c}}^2}{\mu _{\text{micro}}} \, P_{21}^{(3)}(x_{2}),
\\[3mm]
P_{12}''(x_{2})   &= 
- \left(\dfrac{4 \mu_{\text{c}} \mu_{\text{e}}}{\left(\mu_{\text{c}}+\mu_{\text{e}}\right)\mu _{\text{micro}}} + 1\right) P_{21}''(x_{2})
+ \dfrac{\mu \, L_{\text{c}}^2}{\mu _{\text{micro}}} \, P_{21}^{(4)}(x_{2}).
\end{split}
\label{eq:P12_P12prime_P12second_MM}
\end{equation}

The following fourth order differential equation in $P_{21}(x_{2})$ is obtained after substituting eqs.~(\ref{eq:P12_P12prime_P12second_MM}) in eq.~(\ref{eq:equiShe2_MM})$_1$:
\begin{equation}
-C_0 \mu_{\text{c}}
- \dfrac{2 \, \mu_{\text{c}} \mu_{\text{e}} }{\mu_{\text{c}}+\mu_{\text{e}}} \, P_{21}(x_2)
+ \mu \, L_{\text{c}}^2 \, \left(\dfrac{2 \, \mu_{\text{c}}\mu_{\text{e}}}{\left(\mu_{\text{c}} + \mu_{\text{e}}\right) \mu _{\text{micro}} } + 1\right) \, P_{21}''(x_2) 
- \dfrac{\mu ^2 L_{\text{c}}^4}{2 \mu _{\text{micro}}} \, P_{21}^{(4)}(x_2)
= 0 \, ,
\label{eq:equiShe3_MM}
\end{equation}
whose solution is
\begin{align}
P_{21}(x_{2}) =& \,
C_1 e^{-\dfrac{f_1 x_{2}}{L_{\text{c}}}}
+ C_2 e^{ \dfrac{f_1 x_{2}}{L_{\text{c}}}} 
+ C_3 e^{-\dfrac{f_2 x_{2}}{L_{\text{c}}}}
+ C_4 e^{ \dfrac{f_2 x_{2}}{L_{\text{c}}}}
- C_0 f_3 \, ,
\qquad
f_3 \coloneqq \dfrac{\mu _{\text{c}} + \mu _{\text{e}}}{2 \mu _{\text{e}}} \, ,
\notag
\\*[0.5cm]
f_1 \coloneqq& \, \sqrt{
	\dfrac{\mu_{\text{c}} \left(2 \mu_{\text{e}}+\mu _{\text{micro}}\right) + \mu_{\text{e}} \mu _{\text{micro}}
		-\sqrt{\mu_{\text{c}}^2 \left(4 \mu_{\text{e}}^2+\mu _{\text{micro}}^2\right)+2 \mu_{\text{c}} \mu_{\text{e}} \mu _{\text{micro}}^2+\mu_{\text{e}}^2 \mu _{\text{micro}}^2}}
	{\mu \left(\mu_{\text{c}}+\mu_{\text{e}}\right)}} \, ,
	\label{eq:P21SolDiff_MM}
\\*[3mm]
f_2 \coloneqq& \, \sqrt{
	\dfrac{\mu_{\text{c}} \left(2 \mu_{\text{e}}+\mu _{\text{micro}}\right) + \mu_{\text{e}} \mu _{\text{micro}}
		+\sqrt{\mu_{\text{c}}^2 \left(4 \mu_{\text{e}}^2+\mu _{\text{micro}}^2\right)+2 \mu_{\text{c}} \mu_{\text{e}} \mu _{\text{micro}}^2+\mu_{\text{e}}^2 \mu _{\text{micro}}^2}}
	{\mu \left(\mu_{\text{c}}+\mu_{\text{e}}\right)}} \, .
\notag
\end{align}

The last step before applying the boundary conditions is to calculate $P_{12}(x_{2})$ and $u_{1}(x_{2})$:
\begin{equation}
\begin{split}
P_{21}(x_{2}) =
- f_4\left(C_1 e^{-\dfrac{f_1 x_2}{L_{\text{c}}}} + C_2 e^{\dfrac{f_1 x_2}{L_{\text{c}}}}\right)
- f_5\left(C_3 e^{-\dfrac{f_2 x_2}{L_{\text{c}}}} + C_4 e^{\dfrac{f_2 x_2}{L_{\text{c}}}}\right) 
+ C_0 f_6,
\\[3mm]
u_{1}(x_2) =
\dfrac{f_7 \, L_{\text{c}}}{f_1}\left( C_1 e^{-\dfrac{f_1 x_2}{L_{\text{c}}}} - C_2 e^{\dfrac{f_1 x_2}{L_{\text{c}}}}\right)
+ \dfrac{f_8 \, L_{\text{c}}}{f_2}\left( C_3 e^{-\dfrac{f_2 x_2}{L_{\text{c}}}} - C_4 e^{\dfrac{f_2 x_2}{L_{\text{c}}}}\right)
+ C_0 x_2+C_6,
\end{split}
\label{eq:P12SolDiff_uSolDiff_MM}
\end{equation}

where $C_6$ is the last constant that has to be determined thanks to the boundary conditions, and 
\begin{equation}
\begin{array}{rclrcl}
f_4 &\coloneqq& 
1 
- \dfrac{\mu \, f_1^2}{\mu _{\text{micro}}} 
+ \dfrac{4 \mu_{\text{c}} \mu_{\text{e}}}{\mu _{\text{micro}}\left(\mu_{\text{c}} + \mu_{\text{e}}\right)},
\quad
&f_7 &\coloneqq&
- \dfrac{\mu \, f_1^2}{\mu _{\text{micro}}} 
+ \dfrac{2 \mu_{\text{c}} \left(2\mu_{\text{e}} + \mu _{\text{micro}}\right)}{\mu _{\text{micro}}\left(\mu_{\text{c}} + \mu_{\text{e}}\right)},
\\[5mm]
f_5 &\coloneqq&
1 
- \dfrac{\mu \, f_2^2}{\mu _{\text{micro}}} 
+ \dfrac{4 \mu_{\text{c}} \mu_{\text{e}}}{\mu _{\text{micro}}\left(\mu_{\text{c}} + \mu_{\text{e}}\right)},
\quad
&f_8 &\coloneqq&
- \dfrac{\mu \, f_2^2}{\mu _{\text{micro}}} 
+ \dfrac{2 \mu_{\text{c}} \left(2\mu_{\text{e}} + \mu _{\text{micro}}\right)}{\mu _{\text{micro}}\left(\mu_{\text{c}} + \mu_{\text{e}}\right)},
\\[5mm]
f_6 &\coloneqq&
\dfrac{\left(\mu_{\text{c}} + \mu_{\text{e}}\right)\left(2\mu_{\text{e}} + \mu _{\text{micro}}\right)}{2\mu_{\text{e}} \mu _{\text{micro}}}.
\end{array}
\end{equation}

\begin{figure}[H]
	\centering
	\includegraphics[width=0.75\textwidth]{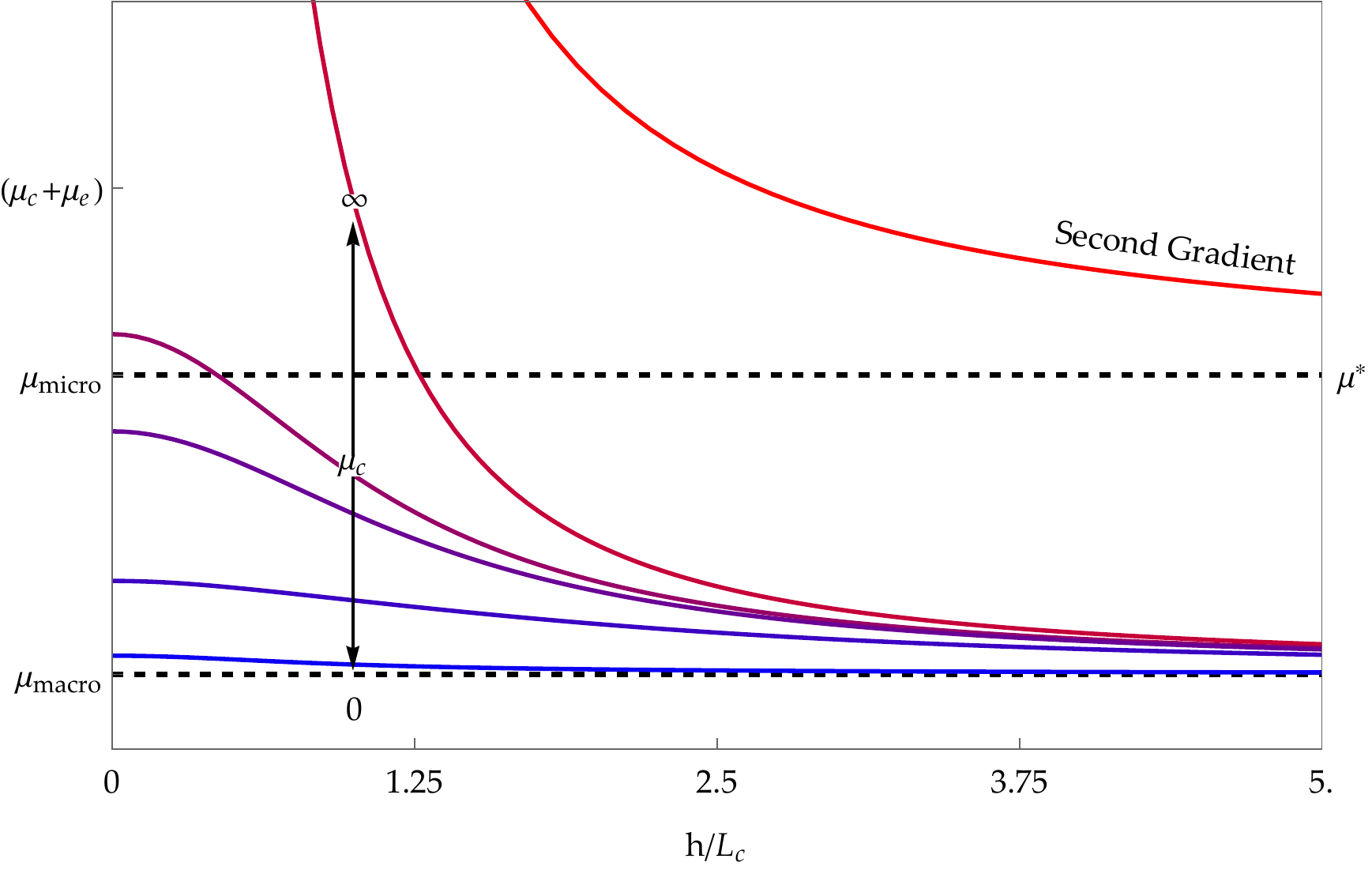}
	\caption{The classical micromorphic model without mixed terms. Plot of the shear stiffness as a function of $L_{\text{c}}$. 
		For small $L_{\text{c}}$ ($h \to \infty$), the shear stiffness is given by $\mu _{\text{macro}}$, where $\mu _{\text{macro}} = \frac{\mu _{\text{micro}} \, \mu_{\text{e}}}{\mu _{\text{micro}} + \mu_{\text{e}}}$. For large $L_{\text{c}}$ ($h \to 0$) the shear stiffness is given by $\mu_{\text{e}} + \mu_{\text{c}}$. If either $\mu_{\text{e}}$ or $\mu_{\text{c}} \to \infty$, there is a singularity appearing in the shear stiffness for $L_{\text{c}} \to \infty$ ($h \to 0$). The values of the elastic parameters that have been used are: $\mu = 1$, $\mu _{\text{e}} = 1.25$, $\mu _{\text{micro}} = 5$, and $\mu _{\text{c}} = \left\{1000,4.3,3,1,0.00001\right\}$.}
	\label{fig:stiff_MM}
\end{figure}

The expressions of the constant of integration, of the solution, and of the measure of the apparent stiffness are too complicated to be reported here.

Nevertheless it is possible to plot how the apparent stiffness behaves while changing $\mu_{\text{c}}$ and $L_{\text{c}}$ (see Fig.~\ref{fig:stiff_MM}).

\section{Simple shear for Forest's micro-strain model with mixed terms}
The micro-strain model considers a symmetric micro-distortion $\boldsymbol{P}=\boldsymbol{S}\in \mbox{Sym}(3)$ and mixed terms like $\langle \boldsymbol{S}, \mbox{sym}\boldsymbol{\nabla u} -\boldsymbol{S}\rangle$, etc.
The expression of the isotropic strain energy for Forest's micro-strain continuum \cite{forest2006nonlinear,hutter2015micromorphic} with (3+6=9) dof's but with just one elastic curvature parameter is:
\begin{equation}
\begin{split}
W \left(\boldsymbol{\nabla u}, \boldsymbol{S}, \boldsymbol{\nabla S}\right) = &
 \, \mu _{\text{e}} \left\lVert \mbox{sym} \boldsymbol{\nabla u} - \boldsymbol{S} \right\rVert^{2}
+ \dfrac{\lambda_{\text{e}}}{2} \mbox{tr}^2 \left(\boldsymbol{\nabla u} - \boldsymbol{S} \right)
+ \mu_{\text{micro}} \left\lVert \boldsymbol{S} \right\rVert^{2} 
+ \dfrac{\lambda_{\text{micro}}}{2} \mbox{tr}^2 \left(\boldsymbol{S} \right)\\[3mm]
& 
+ 2\mu_{\tiny \mbox{mix}} \left\langle \boldsymbol{S},\mbox{sym} \boldsymbol{\nabla u} - \boldsymbol{S} \right\rangle
+ \lambda_{\tiny \mbox{mix}} \mbox{tr} \left(\boldsymbol{S} \right)\mbox{tr} \left(\boldsymbol{\nabla u} - \boldsymbol{S} \right)
+ \dfrac{\mu \, L_{\text{c}}^2}{2} \, \left\lVert \boldsymbol{\nabla S}\right\rVert^2,
\end{split}
\label{eq:energy_FR}
\end{equation}
where $\boldsymbol{S} \in \mbox{Sym}(3)$. The equilibrium equations without body forces are the following:
\begin{equation}
\begin{array}{rrr}
\mbox{Div}\overbrace{\left[
	  2\mu _{\text{e}} \left(\mbox{sym} \boldsymbol{\nabla u} - \boldsymbol{S}\right) + \lambda_{\text{e}} \mbox{tr} \left(\boldsymbol{\nabla u} - \boldsymbol{S} \right) \boldsymbol{\mathbbm{1}}
	+ 2\mu_{\tiny \mbox{mixed}} \, \boldsymbol{S}
	+ \lambda_{\tiny \mbox{mixed}} \mbox{tr} \left(\boldsymbol{S} \right) \boldsymbol{\mathbbm{1}}
	\right]}^{\mathlarger{\widetilde{\boldsymbol{\sigma}}}}
&= \boldsymbol{0},
\\[3mm]
	  2\mu _{\text{e}} \left(\mbox{sym} \boldsymbol{\nabla u} - \boldsymbol{S}\right)
	+ \lambda_{\text{e}} \mbox{tr} \left(\boldsymbol{\nabla u} - \boldsymbol{S} \right) \boldsymbol{\mathbbm{1}}
	- 2\left(\mu_{\text{micro}} - 2\mu_{\tiny \mbox{mixed}}\right) \boldsymbol{S}
	- \left(\lambda_{\text{micro}} - 2\lambda_{\tiny \mbox{mixed}}\right) \mbox{tr} \left(\boldsymbol{S}\right) \boldsymbol{\mathbbm{1}} \, \, &
	\\[3mm]
	- 2\mu_{\tiny \mbox{mixed}} \, \mbox{sym} \boldsymbol{\nabla u}
	- \lambda_{\tiny \mbox{mixed}} \, \mbox{tr} \left(\boldsymbol{\nabla u} \right) \boldsymbol{\mathbbm{1}}
	+ \, \mu \, L_{\text{c}}^{2}\,\mbox{Div} \, \left[\boldsymbol{\nabla S}\right] &= \boldsymbol{0}.
\end{array}
\label{eq:equiMic_FR}
\end{equation}
It is worth noticing the relation of this model to the relaxed micromorphic models: the curvature measure is based on $\boldsymbol{\nabla} \mbox{sym} \boldsymbol{P}$ instead of $\mbox{Curl} \boldsymbol{P}$, $\mu_{\text{c}} \equiv 0$ and the relaxed micromorphic model does not feature mixed terms.
Such a model has also been proposed by \cite{shaat2018reduced} under the name ``reduced micromorphic model'' (``RMM'') \cite{shaat2018reduced}.
It is also worth to highlight that, for the simple shear, the positive definiteness of the material is guaranteed when the following relations hold
\begin{equation}
\mu_{\text{e}} > 0 \, ,
\quad
\mu _{\text{micro}} > 0 \, ,
\quad
\mu_{\text{mix}}^2 < \mu_ e \, \mu _{\text{micro}}.
\end{equation}

No assumptions are made on the components of $\boldsymbol{u}$ and $\boldsymbol{S}$ besides that all the components with index 3 are zero and that the non zero depend only on $x_{2}$:
\begin{equation}
\boldsymbol{u} = \left(
\begin{array}{c}
u_{1}(x_{2}) \\
u_{2}(x_{2}) \\
0 \\
\end{array}
\right),
\quad
\boldsymbol{S} = \left(
\begin{array}{ccc}
S_{11}(x_{2}) & S_{12}(x_{2}) & 0 \\
S_{12}(x_{2}) & S_{22}(x_{2}) & 0 \\
0 & 0 & 0 \\
\end{array}
\right).
\label{eq:non_zero_compo_FR}
\end{equation}

The boundary condition for the simple shear are the following:
\begin{equation}
\begin{array}{rrrlrrrlrrrr}
u_{1}(x_{2} = 0) = 0 \, , &  u_{1}(x_{2} = h) = \boldsymbol{\gamma} \, h \, , &
u_{2}(x_{2} = 0) = 0 \, , &  u_{2}(x_{2} = h) = 0 \, , \\[3mm]
&\boldsymbol{S}(x_{2} = 0) = 0 \, , \;\;\: & \boldsymbol{S}(x_{2} = h) = 0\,. &
\end{array}
\label{eq:BC_FR}
\end{equation}
It is important to underline that the constraint is on all the components of $\boldsymbol{S}$.

After substituting the expressions eq.~(\ref{eq:non_zero_compo_FR}) in eq.~(\ref{eq:equiMic_FR}), the non-trivial equilibrium equations reduces to the following two sets
\begin{align}
\begin{split}
2 \left(\mu_{\text{mix}}-\mu_{\text{e}}\right) S_{12}'(x_2) + \mu_{\text{e}} \, u_1''(x_2) &= 0 \, ,\\
2 \left(\mu_{\text{e}}+\mu _{\text{micro}}-2 \mu_{\text{mix}}\right) S_{12}(x_2) - \left(\mu_{\text{e}} - \mu _{\mbox{\tiny mix}}\right) \, u_1'(x_2) - \mu \, L_{\text{c}}^2 \, S_{12}''(x_2) &= 0 \, ,
\end{split}
\label{eq:equiShe_FR}
\\[3mm]
\begin{split}
\left(-\lambda_{\text{e}}-2 \mu_{\text{e}}+\lambda_{\text{mix}}+2 \mu_{\text{mix}}\right) S_{22}'(x_2) + \left(\lambda_{\text{mix}}-\lambda_{\text{e}}\right) S_{11}'(x_2) + \left(\lambda_{\text{e}}+2 \mu_{\text{e}}\right) u_2''(x_2) &= 0 \, ,\\
\left(\lambda_{\text{e}}+2 \mu_{\text{e}}+\lambda_{\text{micro}}+2 \mu _{\text{micro}}-2 \lambda_{\text{mix}}-4 \mu_{\text{mix}}\right) S_{11}(x_2) + \left(\lambda_{\text{e}}+\lambda_{\text{micro}}-2 \lambda_{\text{mix}}\right) S_{22}(x_2) \qquad&\\
-\left(\lambda_{\text{e}} - \lambda _{\mbox{\tiny mix}}\right) u_2'(x_2) - \mu \, L_{\text{c}}^2 \, S_{11}''(x_2) &= 0 \, ,\\
\left(\lambda_{\text{e}}+2 \mu_{\text{e}}+\lambda_{\text{micro}}+2 \mu _{\text{micro}}-2 \lambda_{\text{mix}}-4 \mu_{\text{mix}}\right) S_{22}(x_2) + \left(\lambda_{\text{e}}+\lambda_{\text{micro}}-2 \lambda_{\text{mix}}\right) S_{11}(x_2) \qquad&\\
-\left(\lambda_{\text{e}} + 2 \mu_{\text{e}} - \lambda_{\text{mix}} - 2\mu_{\text{mix}}\right) u_2'(x_2) - \mu \, L_{\text{c}}^2 \, S_{22}''(x_2) &= 0 \, ,\\
\left(S_{11}(x_2)+S_{22}(x_2)\right) \left(\lambda_{\text{e}}+\lambda_{\text{micro}}-2 \lambda_{\text{mix}}\right) - \left(\lambda_{\text{e}} - \lambda _{\mbox{\tiny mix}}\right) u_2'(x_2) &= 0 \, .
\end{split}
\label{eq:equiSheOther_FR}
\end{align}
It can be noticed that the first set of equations is uncoupled form the second one: the first depends on $S_{12}(x_2)$ and $u_{1}(x_2)$, while the second on $S_{11}(x_2)$, $S_{22}(x_2)$, and $u_{2}(x_2)$.

From the eq.~(\ref{eq:equiSheOther_FR})$_4$ is it possible to evaluate $u_{2}'(x_2)$ and consequently $u_{2}''(x_2)$
\begin{equation}
\begin{array}{l}
u_{2}'(x_2) = 
\dfrac{\left(\lambda_{\text{e}}+\lambda_{\text{micro}}-2\lambda_{\text{mix}}\right)\left(S_{11}\left(x_2\right) + S_{22}\left(x_2\right)\right)}{\lambda_{\text{e}}},
\\[3mm]
u_{2}''(x_2) = 
\dfrac{\left(\lambda_{\text{e}}+\lambda_{\text{micro}}-2\lambda_{\text{mix}}\right)\left(S_{11}'\left(x_2\right) + S_{22}'\left(x_2\right)\right)}{\lambda_{\text{e}}}.
\end{array}
\label{eq:u2prime_FR}
\end{equation}
Substituting eq.~(\ref{eq:u2prime_FR}) in eq.~(\ref{eq:equiSheOther_FR})$_2$ gives a second order differential equation in $S_{11}(x_2)$ which, due to the boundary conditions eqs.~(\ref{eq:BC_FR}), requires $S_{11}(x_2)$ to be identically zero.

After putting to zero $S_{11}(x_2)$ and its derivative eq.~(\ref{eq:equiSheOther_FR})$_3$ becomes a second order differential equation in $S_{22}(x_2)$ which, due to the boundary conditions eqs.~(\ref{eq:BC_FR}), requires $S_{22}(x_2)$ to be identically zero.

It is now clear that $u_{2}(x_2)$ has to be zero due to the fact that $S_{11}(x_2)$ and $S_{22}(x_2)$ are zero, eq.~(\ref{eq:u2prime_FR}) and the boundary conditions eqs.~(\ref{eq:BC_FR}).

\begin{figure}[H]
	\centering
	\includegraphics[width=0.75\textwidth]{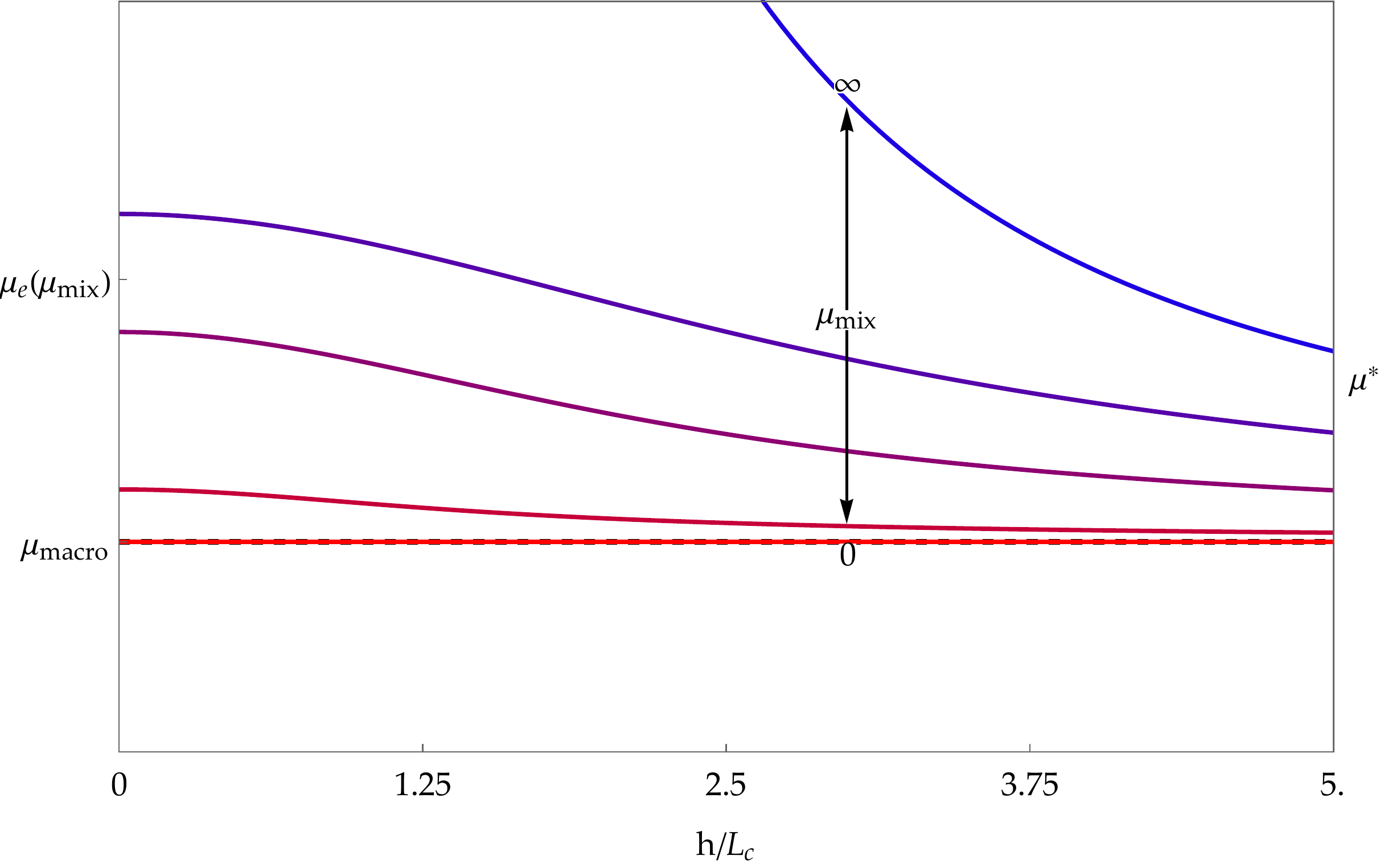}
	\caption{The size-dependent shear-stiffness $\mu^{*}$ for Forest's micro-strain model. The values of the elastic parameters that have been used are: $\mu = 1$, $\mu _{\text{e}} = \left\{1,1.25,2,2.56,21.25\right\}$, $\mu _{\text{micro}} = 5$, and $\mu_{\text{mix}} = \left\{1,2,3,3.5,10\right\}$.}
	\label{fig:stiff_FR}
\end{figure}

After applying the boundary conditions eqs.~(\ref{eq:BC_FR}), the solution of the set of equations eqs.~(\ref{eq:equiShe_FR}) for $u_{1}(x_2)$ and $S_{12}(x_{2})$ is the following:
\begin{equation}
\begin{split}
u_{1}(x_2) &=
\dfrac{\dfrac{f_3 L_{\text{c}}}{f_1 h} \left(\sinh \left(\dfrac{f_1 (h-2 x_2)}{2 L_{\text{c}}}\right)-\sinh \left(\dfrac{f_1 h}{2 L_{\text{c}}}\right)\right) + \dfrac{x_2}{h}\cosh \left(\dfrac{f_1 h}{2 L_{\text{c}}}\right)}{\cosh \left(\dfrac{f_1 h}{2 L_{\text{c}}}\right) - \dfrac{2 f_3 L_{\text{c}}}{f_1 h} \sinh \left(\dfrac{f_1 h}{2 L_{\text{c}}}\right)}\boldsymbol{\gamma}  h,
\\[3mm]
S_{12}(x_{2}) &=
\dfrac{\cosh \left(\dfrac{f_1 h}{2 L_{\text{c}}}\right)-\cosh \left(\dfrac{f_1 (h-2 x_2)}{2 L_{\text{c}}}\right)}{2 \cosh \left(\dfrac{f_1 h}{2 L_{\text{c}}}\right)-\dfrac{4 f_3}{f_1}\dfrac{L_{\text{c}}}{h} \sinh \left(\dfrac{f_1 h}{2 L_{\text{c}}}\right)} \dfrac{\mu_{\text{e}}-\mu_{\text{mix}}}{\mu_{\text{e}}+\mu _{\text{micro}}-2 \mu_{\text{mix}}} \boldsymbol{\gamma},
\\[3mm]
f_1 &\coloneqq \sqrt{\frac{2\left(\mu_{\text{e}} \mu _{\text{micro}}-\mu_{\text{mix}}^2\right)}{\mu \,  \mu_{\text{e}}}}\, ,
\qquad 
f_3 \coloneqq \frac{\left(\mu_{\text{e}}-\mu_{\text{mix}}\right){}^2}{\mu_{\text{e}} \left(\mu_{\text{e}}+\mu _{\text{micro}}-2 \mu_{\text{mix}}\right)}.
\end{split}
\label{eq:S12SolDiff_uSolDiff_FR}
\end{equation}
The following relation is a measure of the higher order stiffness (see Fig.~\ref{fig:stiff_FR})
\begin{equation}
\mu^{*} = 
\dfrac{1}{1 - \dfrac{2 f_3}{f_1} \dfrac{L_{\text{c}}}{h} \tanh \left(\dfrac{f_1 h}{2 L_{\text{c}}}\right)}
\dfrac{\mu_{\text{e}} \mu _{\text{micro}}-\mu_{\text{mix}}^2}{\mu_{\text{e}}+\mu _{\text{micro}}-2 \mu_{\text{mix}}}.
\label{eq:stiff_FR}
\end{equation}

Since the limit $\mu^{*} \rvert _{L_{\text{c}} \to 0} = \dfrac{\mu_{\text{e}} \, \mu _{\text{micro}} - \mu_{\text{mix}}^2}{\mu_{\text{e}} + \mu _{\text{micro}} - 2 \mu_{\text{mix}}} = \mu _{\text{macro}}$ (for $\mu_{\text{mix}} = 0$ the classic value is retrieved) it is possible to evaluate $\mu_{\text{e}}$ with respect to the other constant:

\begin{equation}
\mu _{\text{e}} = 
\dfrac{\mu _{\text{micro}} \, \mu _{\text{macro}} - 2 \mu _{\text{macro}} \, \mu_{\text{mix}} + \mu_{\text{mix}}^2}{\mu _{\text{micro}} - \mu _{\text{macro}}}.
\label{eq:mu_e_FR}
\end{equation}
If $\mu _{\text{macro}}$ is finite, if $\mu_{\text{mix}} \to \infty$ we then have $\mu_{\text{e}}\to \infty$ too.
It is possible decide to keep $\mu_{\text{e}}$ finite, but this will imply that if $\mu_{\text{mix}} \to \infty$, then $\mu _{\text{macro}} \to \infty$ too, which is not desirable.

\section{Simple shear for the second gradient continuum}
The expression of the strain energy for the isotropic second gradient continuum is:
\begin{equation}
W \left(\boldsymbol{\nabla u}, \boldsymbol{\nabla^2 u}\right) = 
  \mu_{\text{micro}} \left\lVert \mbox{sym} \,\boldsymbol{\nabla u} \right\rVert^{2}
+ \dfrac{\lambda_{\text{micro}}}{2} \mbox{tr}^2 \left(\boldsymbol{\nabla u} \right) 
+ \dfrac{\mu \, L_{\text{c}}^2}{2} \left\lVert \boldsymbol{\nabla^2 u}\right\rVert^2,
\label{eq:energy_SGM}
\end{equation}
while the equilibrium equations without body forces are the following:
\begin{equation}
\mbox{Div}\left[\dfrac{}{}
2 \mu _{\text{micro}}\,\mbox{sym}\,\boldsymbol{\nabla u} - \lambda_{\text{micro}} \mbox{tr} \left(\boldsymbol{\nabla u}\right) \boldsymbol{\mathbbm{1}}
+ \mu \, L_{\text{c}}^{2}\,\mbox{Div} \, \left[\boldsymbol{\nabla^2 u} \right]
\right] = \boldsymbol{0}.
\label{eq:equiMic_SGM}
\end{equation}
No assumptions are made on the structure of $\boldsymbol{u}$ besides that the components with index 3 is zero and that the non zero ones depend only on $x_{2}$: $\boldsymbol{u} = \left(u_{1}(x_{2}), u_{2}(x_{2}) ,0\right)^T$.
The boundary condition for the simple shear are the following:
\begin{equation}
\begin{array}{rrrrrrrr}
u_{1}(x_{2} = 0) = 0 \, ,  & u_{1}(x_{2} = h) = \boldsymbol{\gamma} \, h \, , &
u_{2}(x_{2} = 0) = 0 \, ,  & u_{2}(x_{2} = h) = 0 \, , \\[3mm]
u'_{1}(x_{2} = 0) = 0 \, , & u'_{1}(x_{2} = h) = 0 \, , \;\;\: &
u'_{2}(x_{2} = 0) = 0 \, , & u'_{2}(x_{2} = h) = 0 \, . 
\end{array}
\label{eq:BC_SGM}
\end{equation}

After substituting the expression of the displacement field in eq.~(\ref{eq:equiMic_SGM}), the non-trivial equilibrium equations reduces to the following two
\begin{equation}
\mu _{\text{micro}} u_{1}''(x_{2})-\mu  L_{\text{c}}^2 u_{1}^{(4)}(x_{2}) = 0 \, ,
\quad
\left(\lambda_{\text{micro}}+2 \mu _{\text{micro}}\right) u_{2}''(x_{2})-\mu  L_{\text{c}}^2 u_{2}^{(4)}(x_{2}) = 0 \, .
\label{eq:equiShe_Other_SGM}
\end{equation}
It can be noticed that the two fourth order differential equations are uncoupled: the first depends on $u_{1}(x_2)$, while the second on $u_{2}(x_2)$.

\begin{figure}[H]
	\centering
	\includegraphics[width=0.69\textwidth]{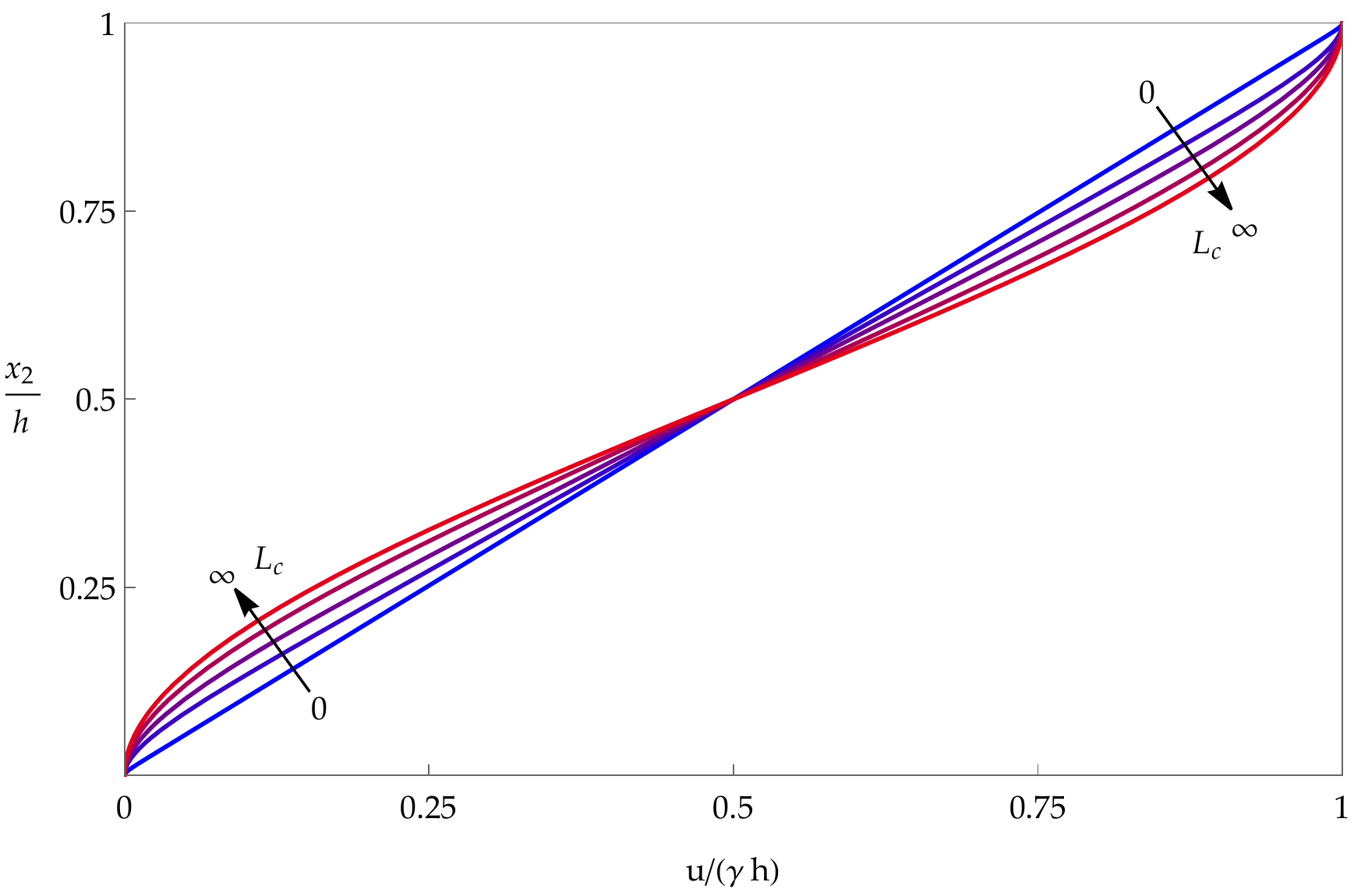}
	\caption{Profile of the dimensionless displacement field for the second gradient model for $f_1 = 2.236$ and different values of $L_{\text{c}} = \left\{0.01, 0.1, 0.2, 0.4, 1000 \right\}$. Note that the deviation from the linear distribution due to the boundary conditions is maximal for $L_{\text{c}} \to \infty$ (also equivalent to $h \to 0$).}
	\label{fig:disp_SGM}
\end{figure}

After applying the boundary conditions to the solution of eqs.~(\ref{eq:equiShe_Other_SGM})$_2$, it results that $u_{2}(x_2)$ has to be zero.
After applying the boundary conditions to the solution of eqs.~(\ref{eq:equiShe_Other_SGM})$_1$, it results that $u_{1}(x_2)$ is:
\begin{equation}
u_{1}(x_{2}) = 
\dfrac{\dfrac{L_{\text{c}} }{f_1 h} \left(\sinh \left(\dfrac{f_1 (h-2 x_2)}{2 L_{\text{c}}}\right)-\sinh \left(\dfrac{f_1 h}{2 L_{\text{c}}}\right)\right) + \dfrac{x_2}{h} \cosh \left(\dfrac{f_1 h}{2 L_{\text{c}}}\right)}{\cosh \left(\dfrac{f_1 h}{2 L_{\text{c}}}\right)-\dfrac{2 L_{\text{c}}}{f_1 h} \sinh \left(\dfrac{f_1 h}{2 L_{\text{c}}}\right)}\boldsymbol{\gamma} \, h \, ,
\qquad
f_1 \coloneqq \sqrt{\dfrac{\mu _{\text{micro}}}{\mu}} \, .
\label{eq:uprime_usecond_SGM}
\end{equation}
A plot of the displacement  file while varying $L_{\text{c}}$ is shown in Fig.~\ref{fig:disp_SGM}.
The following relation is a measure of the higher order stiffness 
\begin{equation}
\mu^{*} = 
\dfrac{1}{1-\dfrac{2 L_{\text{c}}}{f_1 h} \tanh \left(\dfrac{f_1 h}{2 L_{\text{c}}}\right)} \, \mu _{\text{micro}}.
\label{eq:stiff_SGM}
\end{equation}
which is shown in the Section \ref{sec:MM} in Fig.~\ref{fig:stiff_MM}.
Note that here, the size-independent macroscopic shear stiffness $\mu^{*}$ for $L_{\text{c}} \to 0$ ($h \to \infty$) is given bu $\mu _{\text{micro}}$.


\section{Limit cases for the relaxed micromorphic continuum}
In this section we will discuss the limit cases for the relaxed micromorphic model and in particular how it behaves for $0 \gets L_{\text{c}} \to \infty$, and how to retrieve from it the Cosserat model (Sec. \ref{sec:lim_Cos}) and the indeterminate couple stress model (Sec. \ref{sec:lim_IndCoupStress}) as special cases.
\subsection{Limit case for $\mu _{\text{e}} \to \infty$}
This limit implies that $\mbox{sym}\,\boldsymbol{P} = \mbox{sym}\,\boldsymbol{\nabla u}$ and the classic linear elastic solution at the micro-scale ($\mu _{\text{macro}} = \mu _{\text{micro}} \, \mu _{\text{e}} /(\mu _{\text{micro}} +\mu _{\text{e}}) = \mu _{\text{micro}}$) is retrieved:
\begin{equation}
\eval{u_{1}}_{\mu _{\text{e}}\to\infty} = \boldsymbol{\gamma} \, x_{2},
\quad
\eval{\mu^{*}}_{\mu _{\text{e}}\to\infty} = 
\mu_{\text{micro}}
\end{equation}
\subsection{Limit case for $\mu _{\text{c}}$}
\subsubsection{$\mu _{\text{c}} \to 0$}
The classic linear elastic solution at the macro-scale is retrieved:
\begin{equation}
\eval{u_{1}}_{\mu _{\text{c}}\to 0} = \boldsymbol{\gamma} \, x_{2},
\quad
\eval{\mu^{*}}_{\mu _{\text{c}}\to 0} = 
\dfrac{\mu _{\text{micro}}\mu _{\text{e}}}{\mu _{\text{micro}} + \mu _{\text{e}}} =
\mu_{\text{macro}}
\end{equation}
\subsubsection{$\mu _{\text{c}} \to \infty$}
This limit implies that $\mbox{skew}\,\boldsymbol{P} = \mbox{skew}\,\boldsymbol{\nabla u}$.
The solution has identical structure to eqs.~(\ref{eq:solu_BC}) in which $f_1$ and $f_2$ are replaced with their limits ($\widehat{f}_1 = \sqrt{\mu _{\text{e}}/\mu}$ and  $\widehat{f}_2 = \mu _{\text{micro}}/(\widehat{f}_1(\mu _{\text{e}} + \mu _{\text{micro}})) $).
The following solution is retrieved:
\begin{equation}
\begin{split}
\eval{u_{1}}_{\mu _{\text{c}} \to \infty} &= \dfrac{\dfrac{\widehat{f}_2 L_{\text{c}} }{h}\sinh \left(\dfrac{\widehat{f}_1 (h-2 x_{2})}{L_{\text{c}}}\right)+\dfrac{2 x_{2} }{h}\cosh \left(\dfrac{\widehat{f}_1 h}{L_{\text{c}}}\right)-\dfrac{\widehat{f}_2 L_{\text{c}}}{h}\sinh \left(\dfrac{\widehat{f}_1 h}{L_{\text{c}}}\right)}{\cosh \left(\dfrac{\widehat{f}_1 h}{L_{\text{c}}}\right)-\dfrac{\widehat{f}_2 L_{\text{c}} }{h}\sinh \left(\dfrac{\widehat{f}_1 h}{L_{\text{c}}}\right)} \, \dfrac{\boldsymbol{\gamma}  h}{2} ,
\\[3mm]
\eval{P_{21}}_{\mu _{\text{c}} \to \infty} &=
- \dfrac{\sinh \left(\dfrac{\widehat{f}_1 x_{2}}{L_{\text{c}}}\right) \sinh \left(\dfrac{\widehat{f}_1 (h-x_{2})}{L_{\text{c}}}\right)}{\cosh \left(\dfrac{\widehat{f}_1 h}{L_{\text{c}}}\right)-\dfrac{\widehat{f}_2 L_{\text{c}} }{h} \sinh \left(\dfrac{\widehat{f}_1 h}{L_{\text{c}}}\right)} \, \dfrac{\mu _{\text{micro}} }{\mu_{\text{e}} + \mu _{\text{micro}}}\, 
\boldsymbol{\gamma} ,
\\[3mm]
\eval{P_{12}}_{\mu _{\text{c}} \to \infty} &= 
\dfrac{\cosh \left(\dfrac{\widehat{f}_1 h}{L_{\text{c}}}\right) \mu_{\text{e}}  + \sinh \left(\dfrac{\widehat{f}_1 x_2}{L_{\text{c}}}\right) \sinh \left(\dfrac{\widehat{f}_1 (h-x_2)}{L_{\text{c}}}\right) \mu _{\text{micro}} }{ \cosh \left(\dfrac{\widehat{f}_1 h}{L_{\text{c}}}\right)-\dfrac{\widehat{f}_2 L_{\text{c}}}{h}  \sinh \left(\dfrac{\widehat{f}_1 h}{L_{\text{c}}}\right)} \, \dfrac{1}{\mu_{\text{e}}+\mu _{\text{micro}}}\, 
\boldsymbol{\gamma} .
\end{split}
\label{eq:solu_BC_MUC}
\end{equation}

The following relation is a measure of the higher order stiffness 
\begin{equation}
\eval{\mu^{*}}_{\mu _{\text{c}} \to \infty} = 
\dfrac{1}{1-\dfrac{\widehat{f}_2 L_{\text{c}} }{h}\tanh \left(\dfrac{\widehat{f}_1 h}{L_{\text{c}}}\right)} \dfrac{\mu _{\text{e}} \, \mu _{\text{micro}}}{\mu_ e + \mu _{\text{micro}}}.
\end{equation}
\subsection{Limit case for $\mu _{\text{micro}} \to \infty$ (Cosserat, Sect. \ref{sec:Cos})}
\label{sec:lim_Cos}
This limit implies that $\mbox{sym}\,\boldsymbol{P} = 0$, therfore $\boldsymbol{P}=\boldsymbol{A} \in \mathfrak{so}(3)$.
The solution has a similar structure to eqs.~(\ref{eq:solu_BC}) in which $f_2$ is replaced with its limit ($\widetilde{f}_2 = \mu _{\text{c}}/(f_1(\mu _{\text{e}} + \mu _{\text{c}}))$) while $f_1$ doesn't change.
It is important to highlight that for this limit the solution for a Cosserat continuum is retrieved:
\begin{equation}
\begin{split}
\eval{u_{1}}_{\mu _{\text{micro}} \to \infty} &= \dfrac{\dfrac{\widetilde{f}_2 L_{\text{c}} }{h}\sinh \left(\dfrac{f_1 (h-2 x_{2})}{L_{\text{c}}}\right)+\dfrac{2 x_{2} }{h}\cosh \left(\dfrac{f_1 h}{L_{\text{c}}}\right)-\dfrac{\widetilde{f}_2 L_{\text{c}}}{h}\sinh \left(\dfrac{f_1 h}{L_{\text{c}}}\right)}{\cosh \left(\dfrac{f_1 h}{L_{\text{c}}}\right)-\dfrac{\widetilde{f}_2 L_{\text{c}} }{h}\sinh \left(\dfrac{f_1 h}{L_{\text{c}}}\right)} \dfrac{\boldsymbol{\gamma}  h}{2} ,
\\[3mm]
\eval{P_{21}}_{\mu _{\text{micro}} \to \infty} &=
- \dfrac{\sinh \left(\dfrac{f_1 x_{2}}{L_{\text{c}}}\right) \sinh \left(\dfrac{f_1 (h-x_{2})}{L_{\text{c}}}\right)}{\cosh \left(\dfrac{f_1 h}{L_{\text{c}}}\right)-\dfrac{\widetilde{f}_2 L_{\text{c}} }{h} \sinh \left(\dfrac{f_1 h}{L_{\text{c}}}\right)} \, 
\boldsymbol{\gamma},
\\[3mm]
\eval{P_{12}}_{\mu _{\text{micro}} \to \infty} &=
\dfrac{ \sinh \left(\dfrac{f_1 x_2}{L_{\text{c}}}\right) \sinh \left(\dfrac{f_1 (h-x_2)}{L_{\text{c}}}\right) }{ \cosh \left(\dfrac{f_1 h}{L_{\text{c}}}\right)-\dfrac{\widetilde{f}_2 L_{\text{c}}}{h}  \sinh \left(\dfrac{f_1 h}{L_{\text{c}}}\right)} \, \boldsymbol{\gamma}
= - \eval{P_{21}}_{\mu _{\text{micro}} \to \infty}.
\end{split}
\label{eq:solu_BC_MUm}
\end{equation}

The following relation is a measure of the higher order stiffness 
\begin{equation}
\eval{\mu^{*}}_{\mu _{\text{micro}} \to \infty} = 
\dfrac{1}{1-\dfrac{\widetilde{f}_2 L_{\text{c}} }{h}\tanh \left(\dfrac{f_1 h}{L_{\text{c}}}\right)} \, \mu _{\text{e}}.
\end{equation}
\subsection{Limit case for $\mu _{\text{micro}} \to \infty$ and $\mu _{c} \to \infty$ (indet. couple stress, Sect. \ref{sec:IndCoupStress})}
\label{sec:lim_IndCoupStress}
This limit implies that $\mbox{sym}\,\boldsymbol{P} = 0$ and that $\mbox{skew}\,\boldsymbol{P} = \mbox{skew}\,\boldsymbol{\nabla u}$.
The solution has a similar structure to eqs.~(\ref{eq:solu_BC}) in which $f_1$ and $f_2$ are replaced with their limits ($\overline{f}_1 =1/\overline{f}_2 = \sqrt{\mu _{\text{e}}/\mu}$).
It is important to highlight that for these limits. regardless the order in which they are taken, the solution for an Indeterminate couple stress continuum is retrieved:
\begin{equation}
\begin{split}
\eval{u_{1}}_{\mu _{\text{micro}} \to \infty}^{} &= \dfrac{\dfrac{L_{\text{c}} }{\overline{f}_1 h}\sinh \left(\dfrac{\overline{f}_{1} (h-2 x_{2})}{L_{\text{c}}}\right)+\dfrac{2 x_{2} }{h}\cosh \left(\dfrac{\overline{f}_{1} h}{L_{\text{c}}}\right)-\dfrac{L_{\text{c}}}{\overline{f}_1 h}\sinh \left(\dfrac{\overline{f}_{1} h}{L_{\text{c}}}\right)}{\cosh \left(\dfrac{\overline{f}_{1} h}{L_{\text{c}}}\right)-\dfrac{L_{\text{c}} }{\overline{f}_1 h}\sinh \left(\dfrac{\overline{f}_{1} h}{L_{\text{c}}}\right)} \dfrac{\boldsymbol{\gamma}  h}{2} ,
\\[3mm]
\eval{P_{21}}_{\mu _{\text{micro}} \to \infty} &=
- \dfrac{\sinh \left(\dfrac{\overline{f}_{1} x_{2}}{L_{\text{c}}}\right) \sinh \left(\dfrac{\overline{f}_{1} (h-x_{2})}{L_{\text{c}}}\right)}{\cosh \left(\dfrac{\overline{f}_{1} h}{L_{\text{c}}}\right)-\dfrac{L_{\text{c}} }{\overline{f}_1 h} \sinh \left(\dfrac{\overline{f}_{1} h}{L_{\text{c}}}\right)} \, 
\boldsymbol{\gamma},
\\[3mm]
\eval{P_{12}}_{\mu _{\text{micro}} \to \infty} &=
\dfrac{ \sinh \left(\dfrac{\overline{f}_{1} x_2}{L_{\text{c}}}\right) \sinh \left(\dfrac{\overline{f}_{1} (h-x_2)}{L_{\text{c}}}\right) }{ \cosh \left(\dfrac{\overline{f}_{1} h}{L_{\text{c}}}\right)-\dfrac{L_{\text{c}}}{\overline{f}_1 h}  \sinh \left(\dfrac{\overline{f}_{1} h}{L_{\text{c}}}\right)} \, \boldsymbol{\gamma}
= - \eval{P_{21}}_{\mu _{\text{micro}} \to \infty}.
\end{split}
\label{eq:solu_BC_MUm_Muc}
\end{equation}

The following relation is a measure of the higher order stiffness:
\begin{equation}
\eval{\mu^{*}}_{\mu _{\text{micro}} \to \infty} = 
\dfrac{1}{1-\dfrac{ L_{\text{c}} }{\overline{f}_1 h}\tanh \left(\dfrac{\overline{f}_{1} h}{L_{\text{c}}}\right)} \, \mu _{\text{e}}.
\end{equation}
\subsection{Limit case for the characteristic length $L_{\text{c}}$}
\subsubsection{$L_{\text{c}} \to 0$}
Similarly to the limit for $\mu _{\text{c}} \to 0$, the classic linear elastic solution at the macro scale is retrieved:
\begin{equation}
\eval{u_{1}}_{L_{\text{c}}\to 0} = \boldsymbol{\gamma} \, x_{2},
\qquad
\eval{\mu^{*}}_{L_{\text{c}}\to 0} =
\dfrac{\mu _{\text{micro}}\mu _{\text{e}}}{\mu _{\text{micro}} + \mu _{\text{e}}} =\mu_{\text{macro}}.
\end{equation}
\subsubsection{$L_{\text{c}} \to \infty$}
This limit implies that $\mbox{Curl} \boldsymbol{P} = 0$ which requires that $\boldsymbol{P}$ has to be the gradient of a vector field $\boldsymbol{\zeta}$ (Appendix \ref{Sec:appendixA} for more details).

Differently to the limit for $\mu _{\text{e}} \to \infty$, a classical linear elastic solution in between the micro and the macro scale is retrieved:
\begin{equation}
\eval{u_{1}}_{L_{\text{c}}\to \infty}^{} = \boldsymbol{\gamma} \, x_{2},
\qquad
\eval{\mu^{*}}_{L_{\text{c}}\to \infty} = \dfrac{\left(\mu _{\text{e}} + \mu _{\text{c}}\right)\mu _{\text{micro}}}{\mu _{\text{e}} + \mu _{\text{c}} + \mu _{\text{micro}}} < \mu _{\text{micro}}.
\end{equation}
\section{Discussion}
	From the previous sections it is possible to see how the shear stiffness behaves when the ratio between the thickness and the characteristic length $h/L_{\text{c}}$ and the other elastic parameters tend to infinity or to zero.
	The Cosserat model, the classical micromorphic model (excluding when it collapse to the second gradient model), and the  microstrain model cannot be distinguished qualitatively from the size effect under simple shear, but they exhibit a qualitatively different behaviour for thin specimens under bending \cite{hutter2016application,rueger2019experimental,rizzi2020bendingLong}.
	Contrary to all the other models seen in this work, the relaxed micromorphic model is always bounded, with the only exception for the limit of $\mu _{\text{micro}} \to \infty$ when $h/L_{\text{c}} \to 0$, since the relaxed micromorphic model then degenerates into the Cosserat model which in turn collapses itself into the indeterminate couple stress model.
	The relaxed micromorphic model shows also a good balance between the complexity of its solution and the amount of work required in order to obtain that simple and manageable explicit solution (contrary to the classical micromorphic model and the micro-strain model), the physical reasonableness of having a bounded shear stiffness regardless the value of the ratio $h/L_{\text{c}}$ (contrary to the Cosserat model, the indeterminate couple stress model and, the second gradient model), and the freedom given by having more than one elastic parameter (contrary to the indeterminate couple stress model and the second gradient model).
\section{Summary and conclusions}
In the present paper, closed-form solutions of the simple shear problem have been derived for several isotropic linear-elastic micromorphic models, namely for the relaxed micromorphic continuum, the Cosserat continuum, a fully micromorphic continuum and a micro-strain continuum.

Limiting cases like the strain-gradient continuum and the indeterminate couple-stress continuum are considered. Both of the latter show an unbounded shear stiffness if the height of the shear strip becomes very small compared to the intrinsic length $L_{\text{c}}$, which is a physically questionable prediction. In contrast, the stiffness remains bounded for the unconstrained micromorphic models.

Furthermore, the derived solutions show the individual effect of each of the constitutive parameters on the effective shear stiffness of a thin layer. Consequently, these solutions offer a way to calibrate the constitutive parameters, except the bulk moduli, from a series of respective real or virtual shear experiments with a number of specimens of different height $h$.
In this context it shall be pointed out that the unconstrained models (relaxed micromorphic, Cosserat, micro-strain and full micromorphic) yield qualitatively comparable results, so that a parameter set can be presumably calibrated for each of these models from a series of the aforementioned shear experiments.
However, the predictions of these models will differ for size effects under other loading conditions. It is thus an important future task to find and compare the predictions of these models for other loading conditions, like bending or torsion, and to compare their predictions with respective (real or virtual) experiments and to derive guidelines for a favourable choice among the numerous available micromorphic continuum models.


\begingroup
\setstretch{0.8}
\setlength\bibitemsep{3pt}
\printbibliography
\endgroup


\begin{footnotesize}
\appendix
\section{Appendix A}
\label{Sec:appendixA}
Taking the limit of the energy, eq.~(\ref{eq:energy}), for $L_{\text{c}} \to \infty$,  requires that $ \left\lVert \mbox{Curl}\,\boldsymbol{P} \right\rVert = 0$.
This implies that $\boldsymbol{P} = \nabla \boldsymbol{\zeta}$, for some $\zeta : \Omega \to \mathbb{R}^3$, which means that the energy eq.~(\ref{eq:energy}) now becomes

\begin{equation}
\begin{split}
W \left(\boldsymbol{\nabla u}, \boldsymbol{\nabla \zeta}\right) = &
\, \mu _{\text{e}} \left\lVert \mbox{sym} \left(\boldsymbol{\nabla u} - \boldsymbol{\nabla \zeta} \right) \right\rVert^{2}
+\dfrac{\lambda_{\text{e}}}{2} \mbox{tr}^2 \left(\boldsymbol{\nabla u} - \boldsymbol{P} \right) 
+ \mu _{\text{c}} \left\lVert \mbox{skew} \left(\boldsymbol{\nabla u} - \boldsymbol{\nabla \zeta} \right) \right\rVert^{2}
 \\[3mm]
&  
+ \mu_{\text{micro}} \left\lVert \mbox{sym}\,\boldsymbol{\nabla \zeta} \right\rVert^{2}
+ \dfrac{\lambda_{\text{micro}}}{2} \mbox{tr}^2 \left(\boldsymbol{\nabla \zeta} \right),
\end{split}
\label{eq:energyLimit}
\end{equation}
and that eq.~(\ref{eq:equiMic}) turns into
\begin{equation}
\begin{array}{rr}
\mbox{Div}\overbrace{\left[2\mu _{\text{e}}\,\mbox{sym} \left(\boldsymbol{\nabla u} - \boldsymbol{\nabla \zeta} \right) + \lambda_{\text{e}} \mbox{tr} \left(\boldsymbol{\nabla u} - \boldsymbol{\nabla \zeta} \right) \boldsymbol{\mathbbm{1}}
	+ 2\mu _{\text{c}}\,\mbox{skew} \left(\boldsymbol{\nabla u} - \boldsymbol{\nabla \zeta} \right)\right]}^{\mathlarger{\widetilde{\boldsymbol{\sigma}}}}
&= \boldsymbol{0},
\\[3mm]
\widetilde{\sigma}
- 2 \mu _{\text{micro}}\,\mbox{sym}\,\boldsymbol{\nabla \zeta} - \lambda_{\text{micro}} \mbox{tr} \left(\boldsymbol{\nabla \zeta}\right) \boldsymbol{\mathbbm{1}}
&= \boldsymbol{0},
\end{array}
\label{eq:equiMicLim}
\end{equation}
with this boundary conditions $\nabla\,\boldsymbol{u} \cdot \boldsymbol{\tau} = \nabla \boldsymbol{\zeta} \cdot \boldsymbol{\tau}$.
Given eq.~(\ref{eq:equiMicLim})$_1$, eq.~(\ref{eq:equiMicLim})$_2$ reduces to be
\begin{equation}
\mbox{Div}\left[ 
2 \mu _{\text{micro}}\,\mbox{sym}\,\boldsymbol{\nabla \zeta} + \lambda_{\text{micro}} \mbox{tr} \left(\boldsymbol{\nabla \zeta}\right) \boldsymbol{\mathbbm{1}}
\right] = \boldsymbol{0},
\label{eq:equiMicLim2}
\end{equation}
which, for the simple shear problem with the set of boundary conditions $u_1 \left(x_2 = 0(h) \right) = 0(\boldsymbol{\gamma} \, h)$, is equivalent to
\begin{equation}
	\nabla \boldsymbol{\zeta} = 
	\left(\begin{array}{ccc}
	0 & a & 0 \\
	0 & b & 0 \\
	0 & 0 & 0 \\
	\end{array}\right) \, ,
	\quad
	\nabla \boldsymbol{u} = 
	\left(\begin{array}{ccc}
	0 & \boldsymbol{\gamma} & 0 \\
	0 & 0 & 0 \\
	0 & 0 & 0 \\
	\end{array}\right) \, ,
	\label{eq:zeta_u_Lim}
\end{equation}
where $a$ and $b$ are arbitrary constants. This solution to eqs.~(\ref{eq:equiMicLim}) is therefore \textbf{not} unique.
Inserting $\nabla \boldsymbol{u}$ and $\nabla \boldsymbol{\zeta}$ from eq.~(\ref{eq:zeta_u_Lim}) in eq.~(\ref{eq:energyLimit}) the following energy expression is recovered
\begin{equation}
I \left(a, b\right) = 
\mu _{\text{e}} \left[2\left( \dfrac{\boldsymbol{\gamma}-a}{2} \right)^{2} + b^2\right]
+ \mu _{\text{c}} \left[2\left( \dfrac{\boldsymbol{\gamma}-a}{2} \right)^{2} \right]
+ \mu _{\text{micro}} \left[2\left( \dfrac{a}{2} \right)^{2} + b^2\right],
\label{eq:energySubLim}
\end{equation}
which has to be minimized with respect $a$ and $b$ in order to remove the non-uniqueness of the equilibrium system eqs.~(\ref{eq:equiMicLim}).
It is trivial to see that $b$ has to be equal to zero while the following relation
\begin{equation}
\dfrac{\partial}{\partial a} \left(\mu _{\text{e}} \left( \boldsymbol{\gamma}-a \right)^{2} + \mu _{\text{c}} \left( \boldsymbol{\gamma}-a \right)^{2} + \mu _{\text{micro}} a^{2}\right) = \mu _{\text{e}} \left( a - \boldsymbol{\gamma}  \right) + \mu _{\text{c}} \left( a - \boldsymbol{\gamma} \right) + \mu _{\text{micro}} a = 0
\label{eq:minimiza}
\end{equation}
has to be satisfied.
The solution of eq.~(\ref{eq:minimiza}) is 
$a_{\text{min}} = \dfrac{\mu _{\text{e}} + \mu _{\text{c}}}{\mu _{\text{e}} + \mu _{\text{c}} + \mu_{\text{micro}}} \boldsymbol{\gamma}$.
Finally it is possible to substitute $a_{\text{min}}$ into eq.~(\ref{eq:zeta_u_Lim}) obtaining
\begin{equation}
\nabla \boldsymbol{\zeta} = 
\left(\begin{array}{ccc}
0 & \dfrac{\mu _{\text{e}} + \mu _{\text{c}}}{\mu _{\text{e}} + \mu _{\text{c}} + \mu_{\text{micro}}} \boldsymbol{\gamma} & 0 \\
0 & 0 & 0 \\
0 & 0 & 0 \\
\end{array}\right),
\quad
\nabla \boldsymbol{u} = 
\left(\begin{array}{ccc}
0 & \boldsymbol{\gamma} & 0 \\
0 & 0 & 0 \\
0 & 0 & 0 \\
\end{array}\right).
\label{eq:uLim2}
\end{equation}
The solution eq.~(\ref{eq:uLim2}) both satisfy the equilibrium equations and the minimum energy requirement.

\section{Appendix B: the 3D Curl}
\label{Sec:appendixB}

Given $\boldsymbol{A} \in \mathfrak{so}(3)$ 

\begin{equation}
\boldsymbol{A} = 
\left(
\begin{array}{ccc}
0  & -a_3 &  a_2 \\
a_3 &   0  & -a_1 \\
-a_2 &  a_1 &   0 
\end{array}
\right)
\label{eq:appC_A}
\end{equation}
the operator axl is introduced
\begin{equation}
\mbox{axl}
\left(
\begin{array}{ccc}
0  & -a_3 &  a_2 \\
a_3 &   0  & -a_1 \\
-a_2 &  a_1 &   0 
\end{array}
\right)
\coloneqq
\left(
\begin{array}{ccc}
a_1 \\
a_2 \\
a_3 
\end{array}
\right),
\quad
\boldsymbol{A} \cdot \boldsymbol{v} = \left(\mbox{axl} \boldsymbol{A}\right) \times \boldsymbol{v},
\quad
\forall \boldsymbol{v} \in \mathbb{R}^3
\label{eq:appC_axl_A}
\end{equation}
Given the definition eqs.~(\ref{eq:appC_A})-(\ref{eq:appC_axl_A}), the following identity hold (Nye's relation)
\begin{equation}
- \mbox{Curl} \boldsymbol{A} = 
\left( \boldsymbol{\nabla} \mbox{axl} \boldsymbol{A} \right)^T 
- \mbox{tr}\left[\left( \boldsymbol{\nabla} \mbox{axl} \boldsymbol{A} \right)^T\right] \cdot \boldsymbol{\mathbbm{1}},
\quad
\boldsymbol{\nabla} \mbox{axl} \boldsymbol{A}   = 
- \left(\mbox{Curl} \boldsymbol{A}\right)^T
+ \dfrac{1}{2} \mbox{tr}\left[\left( \mbox{Curl} \boldsymbol{A} \right)^T\right] \cdot \boldsymbol{\mathbbm{1}} \, .
\label{eq:appC_ident}
\end{equation}

If we now have $\boldsymbol{A} = \mbox{skew} \, \boldsymbol{\nabla u}$ it is possible to show that 
\begin{equation}
- \left( \mbox{Curl} \, \mbox{skew} \, \boldsymbol{\nabla u} \right)^T= 
\boldsymbol{\nabla} \mbox{axl}  \left(\mbox{skew} \, \boldsymbol{\nabla u}\right)  ,
\quad
\dfrac{1}{2} \, \mbox{curl} \boldsymbol{u} = 
\mbox{axl}\left( \mbox{skew} \boldsymbol{\nabla u}\right),
\label{eq:nye_2}
\end{equation}
which lead to this identity for the full Curl,
$\left\lVert \mbox{Curl} \, \mbox{skew} \, \boldsymbol{\nabla u} \right\rVert^{2}
=
\dfrac{1}{4} \left\lVert \boldsymbol{\nabla} \mbox{curl} \, \boldsymbol{u} \right\rVert^{2}.
$

It is also highlighted here that, thanks to eq.~(\ref{eq:nye_2})$_1$ this relation holds
$
-  \mbox{Curl} \, \mbox{skew} \, \boldsymbol{\nabla u}  = 
 \mbox{Curl} \, \mbox{sym} \, \boldsymbol{\nabla u}
$
(see \cite{ghiba2017variant}),
which implies that choosing the symmetric or the skew-symmetric part of the gradient of the displacement do not make any difference besides a sign.
\section{Appendix C: the planar Curl}
\label{Sec:appendixC}
Given the following structure of $\boldsymbol{P}$ and convention for the Curl ($\cdot$) operator
\begin{equation}
\boldsymbol{P}
=
\left(
\begin{array}{ccc}
P_{11}\left(x_1,x_2\right) & P_{12}\left(x_1,x_2\right) & 0 \\
P_{21}\left(x_1,x_2\right) & P_{22}\left(x_1,x_2\right) & 0 \\
0 & 0 & 0
\end{array}
\right),
\quad
\mbox{Curl}~\boldsymbol{P}
=
\left(
\begin{array}{c}
\mbox{curl}\left( P_{11} \, , \right. \, P_{12} \, , \, \left. P_{13} \right) \\
\mbox{curl}\left( P_{21} \, , \right. \, P_{22} \, , \, \left. P_{23} \right) \\
\mbox{curl}\left( P_{31} \, , \right. \, P_{32} \, , \, \left. P_{33} \right) 
\end{array}
\right),
\end{equation}
the matrix Curl $\boldsymbol{P}$ becomes
\begin{equation}
\mbox{Curl}~\boldsymbol{P}
=
\left(
\begin{array}{ccc}
0 & 0 & P_{12,1} - P_{11,2} \\
0 & 0 & P_{22,1} - P_{21,2} \\
0 & 0 & 0 \\
\end{array}
\right)
=
\left(
\begin{array}{ccc}
0 & 0 & \alpha \\
0 & 0 & \beta \\
0 & 0 & 0 \\
\end{array}
\right),
\end{equation}
which implies that, since ``dev $ \mbox{Curl}~\boldsymbol{P} = \mbox{Curl}~\boldsymbol{P}$'', ``$\left\lVert \mbox{dev} \, \mbox{sym} \, \mbox{Curl} \, \boldsymbol{P} \right\rVert^2 = \left\lVert \mbox{sym} \, \mbox{Curl} \, \boldsymbol{P} \right\rVert^2 = \left\lVert \mbox{skew} \, \mbox{Curl} \, \boldsymbol{P} \right\rVert^2 = \dfrac{1}{2} \, \left\lVert \mbox{Curl} \, \boldsymbol{P} \right\rVert^2 $'' and ``$\mbox{tr} \, \mbox{Curl} \, \boldsymbol{P} = 0$'', there can be only two relevant curvature parameters.
For the simple shear problem we end up with one curvature parameter, since $P_{11}$,$P_{22}$, and $P_{21,2}$ are zero.

For the simple shear problem, the couple stress model and the ones which are derived from it, $\boldsymbol{P}$ is replaced by skew $\boldsymbol{\nabla u}$, which implies that:
\begin{equation}
\boldsymbol{\nabla u} = 
\left(
\begin{array}{ccc}
0 & u_{1,2}(x_{2}) & 0 \\
0 & u_{2,2}(x_{2}) & 0 \\
0 & 0 & 0
\end{array}
\right),
\quad
\mbox{skew}~\boldsymbol{\nabla u} = 
\left(
\begin{array}{ccc}
0 & \dfrac{u_{1,2}(x_{2})}{2} & 0 \\
-\dfrac{u_{1,2}(x_{2})}{2} & 0 & 0 \\
0 & 0 & 0
\end{array}
\right).
\label{eq:non_zero_compo_Ind_P}
\end{equation}

Accordingly, the expressions ``Curl skew $\boldsymbol{\nabla u}$", the ``sym Curl skew $\boldsymbol{\nabla u}$", and the ``skew Curl skew $\boldsymbol{\nabla u}$" become, 

\begin{equation}
\begin{array}{c}
\mbox{Curl}~\mbox{skew}~\boldsymbol{\nabla u} = 
\left(
\begin{array}{ccc}
0 & 0 & 0 \\
0 & 0 & \dfrac{u_{1,22}(x_{2})}{2} \\
0 & 0 & 0
\end{array}
\right),
\quad
\begin{array}{c}
\mbox{sym}\\
\mbox{skew}
\end{array}
\mbox{Curl}~\mbox{skew}~\boldsymbol{\nabla u} = 
\left(
\begin{array}{ccc}
0 & 0 & 0 \\
0 & 0 & \dfrac{u_{1,22}(x_{2})}{4} \\
0 & \pm \dfrac{u_{1,22}(x_{2})}{4} & 0
\end{array}
\right),
\\[15mm]
\mbox{skew}~\mbox{Curl}~\mbox{skew}~\boldsymbol{\nabla u} = 
\left(
\begin{array}{ccc}
0 & 0 & 0 \\
0 & 0 & \dfrac{u_{1,22}(x_{2})}{4} \\
0 & -\dfrac{u_{1,22}(x_{2})}{2} & 0
\end{array}
\right).
\end{array}
\label{eq:non_zero_compo_Ind_Grad}
\end{equation}
This shows that $\dfrac{1}{2} \left\lVert \mbox{Curl} \, \mbox{skew}\boldsymbol{\nabla u} \right\rVert^2
=
\left\lVert \mbox{sym}  \, \mbox{Curl} \, \mbox{skew}\boldsymbol{\nabla u} \right\rVert^2 
=
\left\lVert \mbox{skew} \, \mbox{Curl} \, \mbox{skew}\boldsymbol{\nabla u} \right\rVert^2 $.
\end{footnotesize}
\end{document}